\documentclass[3p,,preprint,12pt]{elsarticle}
\makeatletter\if@twocolumn\PassOptionsToPackage{switch}{}\else\fi\makeatother

\usepackage{tabulary,xcolor}
\usepackage{amsfonts,amsmath,amssymb}
\usepackage[T1]{fontenc}
\makeatletter
\let\save@ps@pprintTitle\ps@pprintTitle
\def\hlinewd#1{%
  \noalign{\ifnum0=`}\fi\hrule \@height #1%
  \futurelet\reserved@a\@xhline}

\AtBeginDocument{\ifNAT@numbers \biboptions{sort&compress}\fi}
\makeatother

\usepackage{ifluatex}
\ifluatex
\usepackage{fontspec}
\defaultfontfeatures{Ligatures=TeX}
\usepackage[]{unicode-math}
\unimathsetup{math-style=TeX}
\else
\usepackage[utf8]{inputenc}
\fi
\ifluatex\else\usepackage{stmaryrd}\fi
\usepackage{float}
\usepackage{amsfonts}
\usepackage{dsfont}
\usepackage{graphicx}
\usepackage{wrapfig}
\usepackage{lipsum}
\usepackage{amsmath}
\usepackage[usestackEOL]{stackengine}
\usepackage{tabularx}

\usepackage{amsfonts}
\usepackage{dsfont}
\usepackage{lastpage}
\usepackage{fancyhdr}
\usepackage{graphicx}
\usepackage{lipsum} 
\usepackage[usestackEOL]{stackengine}
\usepackage{url,multirow,morefloats,floatflt,cancel,tfrupee}
\makeatletter

\AtBeginDocument{\@ifpackageloaded{textcomp}{}{\usepackage{textcomp}}}
\makeatother
\usepackage{colortbl}
\usepackage{xcolor}
\usepackage{pifont}
\usepackage[nointegrals]{wasysym}

\makeatletter

\def\mcWidth#1{\csname TY@F#1\endcsname+\tabcolsep}

\def\cAlignHack{\rightskip\@flushglue\leftskip\@flushglue\parindent\z@\parfillskip\z@skip}
\def\rAlignHack{\rightskip\z@skip\leftskip\@flushglue \parindent\z@\parfillskip\z@skip}

\@ifundefined{etal}{}{}

\usepackage{ifxetex}
\ifxetex\else\if@twocolumn\@ifpackageloaded{stfloats}{}{\usepackage{dblfloatfix}}\fi\fi

\AtBeginDocument{
\expandafter\ifx\csname eqalign\endcsname\relax
\def\eqalign#1{\null\vcenter{\def\\{\cr}\openup\jot\m@th
  \ialign{\strut$\displaystyle{##}$\hfil&$\displaystyle{{}##}$\hfil
      \crcr#1\crcr}}\,}
\fi
}

\AtBeginDocument{%
  \@ifpackageloaded{endfloat}%
   {\renewcommand\efloat@iwrite[1]{\immediate\expandafter\protected@write\csname efloat@post#1\endcsname{}}}{\newif\ifefloat@tables}%
}%

\def\BreakURLText#1{\@tfor\brk@tempa:=#1\do{\brk@tempa\hskip0pt}}
\let\lt=<
\let\gt=>
\def\processVert{\ifmmode|\else\textbar\fi}

\@ifundefined{subparagraph}{
\def\subparagraph{\@startsection{paragraph}{5}{2\parindent}{0ex plus 0.1ex minus 0.1ex}%
{0ex}{\normalfont\small\itshape}}%
}{}

\newcommand\role[1]{\unskip}
\newcommand\aucollab[1]{\unskip}
\newcommand{\holl}{\texttt{\textsf{HOL Light}}}

\@ifundefined{tsGraphicsScaleX}{\gdef\tsGraphicsScaleX{1}}{}
\@ifundefined{tsGraphicsScaleY}{\gdef\tsGraphicsScaleY{.9}}{}
\def\checkGraphicsWidth{\ifdim\Gin@nat@width>\linewidth
	\tsGraphicsScaleX\linewidth\else\Gin@nat@width\fi}

\def\checkGraphicsHeight{\ifdim\Gin@nat@height>.9\textheight
	\tsGraphicsScaleY\textheight\else\Gin@nat@height\fi}

\def\fixFloatSize#1{}
\let\ts@includegraphics\includegraphics

\def\inlinegraphic[#1]#2{{\edef\@tempa{#1}\edef\baseline@shift{\ifx\@tempa\@empty0\else#1\fi}\edef\tempZ{\the\numexpr(\numexpr(\baseline@shift*\f@size/100))}\protect\raisebox{\tempZ pt}{\ts@includegraphics{#2}}}}

\AtBeginDocument{\def\includegraphics{\@ifnextchar[{\ts@includegraphics}{\ts@includegraphics[width=\checkGraphicsWidth,height=\checkGraphicsHeight,keepaspectratio]}}}

\DeclareMathAlphabet{\mathpzc}{OT1}{pzc}{m}{it}

\def\URL#1#2{\@ifundefined{href}{#2}{\href{#1}{#2}}}

\def\UrlOrds{\do\*\do\-\do\~\do\'\do\"\do\-}%
\g@addto@macro{\UrlBreaks}{\UrlOrds}

\edef\fntEncoding{\f@encoding}

\makeatother

\newif\ifmultipleabstract\multipleabstractfalse%
\newtheorem{definition}{Definition}[section]

\newtheorem{theorem}{Theorem}[section]

\usepackage{lineno}

\begin{document}

\begin{frontmatter}

\title{Formalization of Bond Graph using Higher-order-logic Theorem Proving}

\author[1]{Ujala Qasim}
\ead{uqasim.mscse16@rcms.nust.edu.pk}
\author[2]{Adnan Rashid\corref{cor1}}
\ead{adnan.rashid@seecs.nust.edu.pk}
\author[2]{Osman Hasan}
\ead{osman.hasan@seecs.nust.edu.pk}

\cortext[cor1]{Corresponding author}

\address[1]{{Research Center for Modeling and Simulations (RCMS)} \\
{National University of Sciences and Technology (NUST), Islamabad, Pakistan}
}
\address[2]{{School of Electrical Engineering and Computer Science (SEECS)} \\
{National University of Sciences and Technology (NUST), Islamabad, Pakistan}
}


\begin{abstract}
Bond graph is a unified graphical approach for describing the dynamics of complex engineering and physical systems and is widely adopted in a variety of domains, such as, electrical, mechanical, medical, thermal and fluid mechanics. Traditionally, these dynamics are analyzed using paper-and-pencil proof methods and computer-based techniques. However, both of these techniques suffer from their inherent limitations, such as human-error proneness, approximations of results and enormous computational requirements. Thus, these techniques cannot be trusted for performing the bond graph based dynamical analysis of systems from the safety-critical domains like robotics and medicine. Formal methods, in particular, higher-order-logic theorem proving, can overcome the shortcomings of these traditional methods and provide an accurate analysis of these systems. It has been widely used for analyzing the dynamics of engineering and physical systems.
In this paper, we propose to use higher-order-logic theorem proving for performing the bond graph based analysis of the physical systems. In particular, we provide formalization of bond graph, which mainly includes functions that allow conversion of a bond graph to its corresponding mathematical model (state-space model) and the verification of its various properties, such as, stability. To illustrate the practical effectiveness of our proposed approach, we present the formal stability analysis of a prosthetic mechatronic hand using \holl~theorem prover. Moreover, to help non-experts in HOL, we encode our formally verified stability theorems in MATLAB to perform the stability analysis of an anthropomorphic prosthetic mechatronic hand.
\end{abstract}

\end{frontmatter}
\section{Introduction}\label{SEC:Introduction}

Bond Graph (BG) is a linear, labelled, directed and domain-independent graphical approach for modelling dynamics of physical systems and is widely adopted for capturing the dynamics of physical systems belonging to multidisciplinary energy domains, such as, electromechanical, hydroelectric and mechatronics. The concept of BG was introduced by H.M. Paynter~\cite{paynter1961analysis}, of Massachusetts Institute of Technology (MIT) in 1961, to describe the dynamics of different systems belonging to multidisciplinary domains and exhibiting analogous dynamical behaviour.
Signal-flow graphs and the block diagram representations are the other graphical approaches used for capturing the dynamics of the physical systems. However, BG models are generally preferred due to their ability to communicate seamlessly between different components belonging to multidisciplinary domains based on the bi-directional flow of information and thus provide a deep insight to the computational structure of the physical systems. Due to these distinguishing features, BG have been widely used in automobiles~\cite{wang2010modeling}, biological systems~\cite{diaz2011formalization}, aerospace~\cite{granda2003automated,gawthrop2007bond} and transportation systems~\cite{iordanova2006bond}. The first step in BG based dynamical analysis of a system is to apply the causal equations on different components of  a system. Next step involves solving the set of equations simultaneously to obtain the corresponding set of differential equations. These equations are transformed into state-space representations by applying various properties of vectors and matrices. The final step is based on analyzing various properties, such as, stability, of the state-space model of the underlying system.
\par
Traditionally, the BG based analysis is performed using paper-and-pencil proof method. However, the analysis is prone to error due to the highly involved human manipulation for analyzing the complex physical systems and thus could not ensure absolute accuracy of the analysis. Similarly, computer based symbolic and numerical techniques have been widely used for analyzing BG based models of the systems. Some of the widely used tools are N-PORT (ENPORT)~\cite{BGenport,hales2001enport}, DEscription and SImulation Systems (DESIS)~\cite{delgado1991desis}, Model Transformation Tools (MTT)~\cite{mtt2005tool}, Computer Aided Modeling Program with Graphical Input (CAMP-G)~\cite{granda1997new} and Twente Simulator (20-sim)~\cite{20simtoolbox}. However, these numerical methods involve the approximation of the mathematical expressions and results due to the finite precision of computer arithmetic. Moreover, they involve a finite number of iterations based on limited computational resources and computer memory. Similarly, the symbolic techniques are based on a large numbers of unverified symbolic algorithms present in the core of the associated tools. Based on the above-mentioned limitations, the computer-based methods cannot be trusted for performing the BG based analysis of the safety-critical systems, such as, aerospace, medicine and transportation, where an inaccurate analysis can lead to disastrous consequences.
\par
\begin{figure}[!ht]
	\centering
	\resizebox{0.8\hsize}{0.34\textheight}{\includegraphics{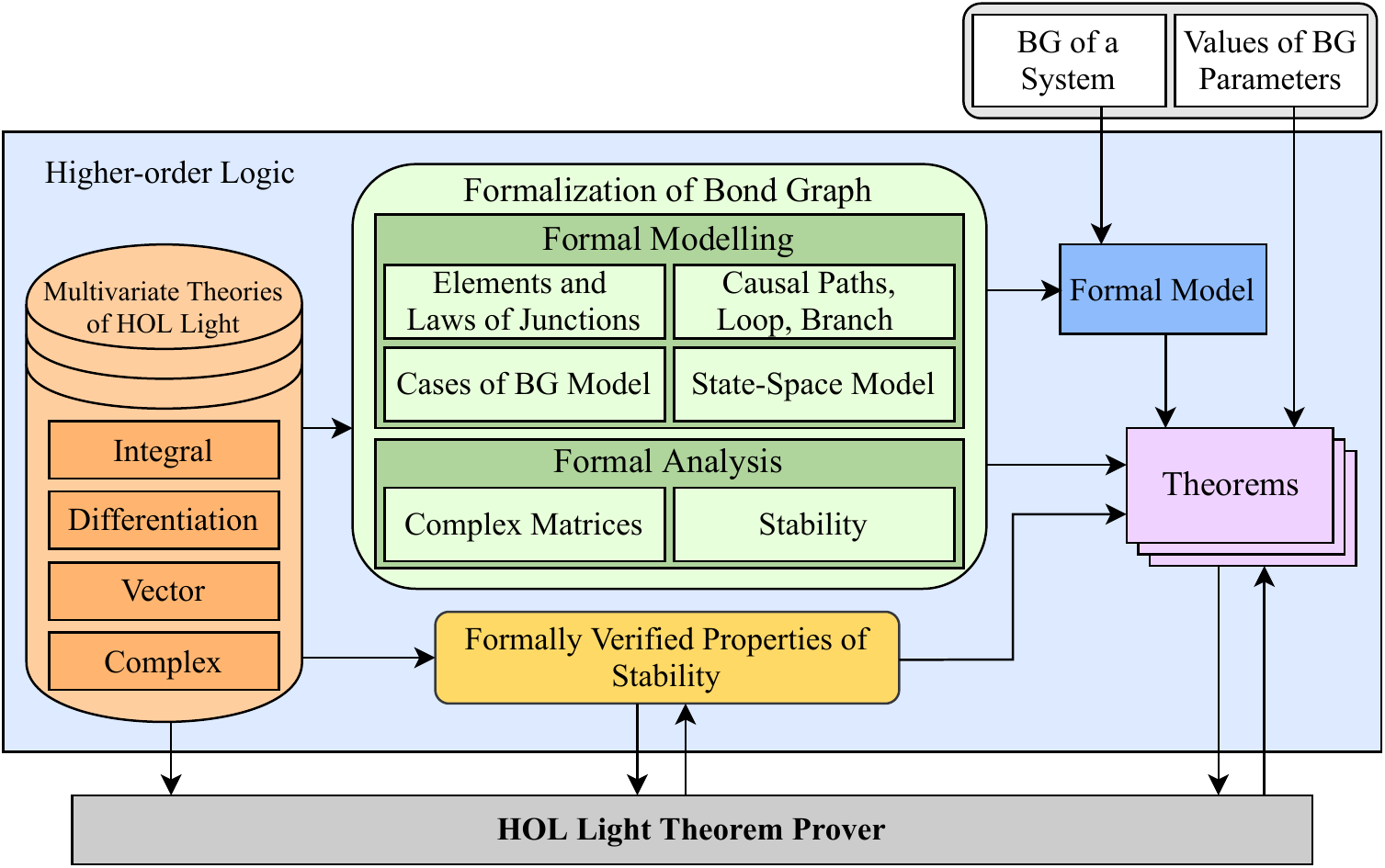}}
          \caption{Proposed Framework}
	\label{fig:PM}
\end{figure}

Formal methods are computer-based mathematical analysis techniques that involve constructing a mathematical model based on an appropriate logic and verification of its various properties based on deductive reasoning. Higher-order-logic theorem proving is a widely adopted formal method for performing an accurate analysis of the engineering and physical systems.
In this paper, we propose a higher-order-logic based framework, as depicted in Figure~\ref{fig:PM}, for the BG based analysis of the physical systems. It mainly involves the formalization of the BG models and the formal verification of their various properties using multivariate calculus theories of \holl~theorem prover~\cite{harrison-hol-light}. To illustrate the practical effectiveness of our proposed framework, we formally analyze the dynamics of an anthropomorphic mechatronic prosthetic hand~\cite{saeed2019comprehensive} based on the formalization of BG.

Our proposed framework accepts a BG representation of a system and its corresponding parameters, such as, the values of its different components from a user. The first step is the development of the mathematical model (state-space representation), which involves a conversion of the BG representation to a set of differential equations. It is based on the application of the laws of junctions and the causality of paths on the BG representation of the underlying system. The development of the state-space model from a set of differential equations is based on the complex-valued vectors and matrices, which are developed as a part of our proposed formalization. The next step is to use the state-space models of the BG representation to formally analyze various dynamical properties, such as, stability, of the underlying system. For a practical  illustration of our proposed formalization, we present a formal stability analysis of an anthropomorphic mechatronic prosthetic hand~\cite{saeed2019comprehensive}. Moreover, to facilitate a non-expert HOL user, we encode our formally verified stability theorems in MATLAB to perform stability analysis of an anthropomorphic prosthetic mechatronic hand.
\par
The rest of the paper is organized as follows: We provide related work in Section~\ref{SEC:relwork} about the formal verification of engineering systems involving complex dynamics. Section \ref{SEC:prelim}~provides preliminaries that include a brief introduction of the \holl~theorem prover, multivariable calculus theories of \holl~and BGs. Section \ref{SEC:flow of bg}~describes the proposed algorithm for the formal analysis of BGs and some set of assumptions/rules for the formal modelling of BGs. Section \ref{SEC:Formalization of BG model}~presents the formalization of the fundamental components of BG representations of different systems. We provide the definition of stability, formally verified general stability theorem and present various properties of stability analysis in Section \ref{SEC:bond_graphs_properties}~of the paper. Section \ref{SEC:app} provides the stability analysis of an anthropomorphic mechatronic prosthetic hand. Finally, Section \ref{SEC:Conclusion} concludes the paper.

\section{Related Work}\label{SEC:relwork}

Traditionally, the BG based analysis of the engineering and physical systems is performed using paper-and-pencil proofs, and computer based symbolic and numerical techniques~\cite{granda2003automated,gawthrop2007bond,iordanova2006bond,wang2010modeling}. However, these methods suffer from their inherent limitations, such as, human error proness, approximation of the mathematical expressions and limited computational resources. Formal methods, in particular, higher-order-logic theorem proving has been widely used for performing an accurate analysis of the engineering and physical and systems. BG based formal analysis using a theorem prover is mainly based on some foundational libraries, such as, vectors and matrices. These vectors and matrices libraries have been formalized in various higher-order-logic theorem provers, such as HOL4, \holl, Coq and Isabelle~\cite{shi2020formalization,harrison-complexanalysis,holli8850640:online,Stand2314151:online,thiemann2016algebraic, Theor8670475:online,Theor4411608:online}. Among these formal libraries, the one available in \holl~is quite comprehensive, especially in the reasoning support for vectors and matrices, and is thus suitable for the proposed formalization of BG.
\par
\holl~has been extensively used for formally analyzing many engineering and physical systems. Rashid et al.~\cite{rashid2016formalization,rashid2019formal,rashid2017formalization,rashid2019formalization} formalized the Laplace and the Fourier transforms using \holl, which are the major mathematical techniques used for analyzing the dynamics of the continuous-time systems. Moreover, the authors used these formalizations for formally analyzing an automobile suspension system, a MEMs accelerometer and a $4-\pi$ soft-error crosstalk model~\cite{rashid2021formal,rashid2018formalization}. Similarly, Rashid et al.~\cite{rashid2017formal,rashid2020formal} formally verified a cell injection system upto $4$-DOF using \holl. The authors verified various coordinate systems and their interrelationship, which are vital for capturing the position and relative movement of a robotic cell injection system. Moreover, they also formally verified the solution of differential equations modeling the continuous dynamics of the system.
Beillahi et al.~\cite{beillahi2016formal} formalized the signal-flow-graph theory using \holl~to perform the formal analysis of engineering systems namely, the PANDA Vernier resonator and the z-source impedance network.
Similarly, Siddique et al.~\cite{siddique2012formal} formalized geometrical optics using \holl~for analyzing
optical and laser systems. The authors formally verified frequently used optical components like thin lens, thick lens and plane parallel plate and performed the stability analysis of Fabry P\'{e}rot resonator and Z-shaped resonator.
Recently, Abed et al. performed the dynamical analysis of Unmanned Aerial Vehicles (UAVs)~\cite{abed2020formaluav} and synthetic biological circuits~\cite{abed2020formal,rashid2020formalsynb}. Similarly, Ahmed et al.~\cite{ahmed2018formal} performed a stability analysis of power converters using \holl. However, none of these contributions provide the BG based analysis of systems, which is the scope of the current paper. A more detailed account of the formal analysis of the engineering and physical systems can be found at~\cite{rashid2019formalsurvey}.
\section{Preliminaries} \label{SEC:prelim}
In this section, we provide an introduction to the \holl~theorem prover, its multivariate calculus theories and BGs which are necessary for understanding rest of the paper.
\subsection{\holl~Theorem Prover} \label{SUBSEC:HOL_Light_theorem_prover}
\holl~is an interactive theorem prover developed by John Harrison in 1996 at the University of Cambridge. It belongs to the family of HOL theorem provers and is used for developing proofs in higher-order logic. \holl~is built using the functional programming language objective CAML (OCaml). It has been extensively used for the formal verification of both hardware~\cite{binyameen2013formal,rashid2017formal,siddique2013formal} and software systems~\cite{schumann2001automated} along with the formalization of mathematics.
\par
A theorem in \holl~is verified by applying the basic axioms and primitive inference rules or any other previously verified theorems/inference rules. Generally, a \holl~theorem is of the form \texttt{\textsf{A1, A2, . . . , An}} $\Rightarrow$ \texttt{\textsf{C}}, where \texttt{\textsf{A1, A2, . . . , An}} model the set of assumptions and \texttt{\textsf{C}} models the conclusion and is the main goal of the \holl~theorem. The proof of a theorem involves the concept of a tactic (an ML function), which divides the main goal into subgoals. These tactics are repeatedly used to reduce or simplify the main goal (required theorem) into intermediate subgoals until they match with the assumptions of the \holl~theorem, concluding the proof of the required theorem.
\par
\holl~provides an extensive support of the formal libraries for the multivariate calculus, such as, differentiation, integration, transcendental and topology, which have been extensively used in the proposed formalization of BGs. The availability of these mathematical theories is one of the main motivations for opting \holl~for our proposed formalization of BGs.
Table~\ref{tab:symbols} provides some of the frequently used \holl~symbols and functions in our formalization.

\begin{table}[h]
\centering
\footnotesize
\renewcommand{\arraystretch}{1.00}
\begin{tabular}{| c | c | c |}
\hline
\textbf{\holl~Symbols} & \textbf{Standard Symbols} & \textbf{Meaning} \\
\hline
\textbf{/\textbackslash} & and & Logical \textit{and}\\
\hline
\textbf{\textbackslash /} & or & Logical \textit{or}\\
\hline
\textbf{$\sim$} & not & Logical \textit{negation}\\
\hline
\textbf{==\textgreater} & \textbf{$\longrightarrow$} & Implication \\
\hline
\textbf{\textless=\textgreater} & \textbf{=} & Equality\\
\hline
\texttt{\textsf{!x.t}} & \textit{$\forall$x.t} & For all \textit{x : t}\\
\hline
\texttt{\textsf{$\lambda$x.t}} & \textit{$\lambda$x.t} & Function that maps \textit{x to t(x)}\\
\hline
\texttt{\textsf{num}} & \{0,1,2,\ldots\}& Positive Integers data type \\
\hline
\texttt{\textsf{real}} & All real numbers & Real data type\\
\hline
\texttt{\textsf{complex}} & All complex numbers& Complex data type\\
\hline
\texttt{\textsf{SUC n}} & \text{(n + 1)} & Successor of natural number\\
\hline
\texttt{\textsf{\&a}} & $\mathbb{N \to R}$ & Typecasting from Integers to Reals \\
\hline
\texttt{\textsf{lift x}} & $\mathbb{R \to}\mathbb{ R}^{1}$ & Map Reals to 1-Dimensional Vectors\\
\hline
\texttt{\textsf{drop x}} & $\mathbb{R}^{1}\mathbb{\to R}$ & Map 1-Dimensional Vectors to Reals\\
\hline
\texttt{\textsf{EL n L}} & \textit{element}& nth element of list L\\
\hline
\texttt{\textsf{Append L1 L2}} & \textit{append} & Combine two lists together\\
\hline
\texttt{\textsf{LENGTH L}} & \textit{length} & Length of list L\\
\hline
\texttt{\textsf{(a,b)}} & \textit{a $\times$ b}& A pair of two elements\\
\hline
\texttt{\textsf{FST}} & \textit{fst(a,b) = a} & First component of a pair\\
\hline
\texttt{\textsf{SND}} & \textit{snd(a,b) = b}& Second component of a pair\\
\hline
\texttt{\textsf{\{x|P(x)\}}} & \textit{\textbraceleft x \textbar P(x)\textbraceright}& Set of all \textit{x} such that \textit{P(x)}\\
\hline
\end{tabular}
\caption{\holl~Symbols and Functions}
\label{tab:symbols}
\end{table}


\subsection{Multivariable Calculus Theories in \holl} \label{SUBSEC:Mult_cal_theories}
In \holl, a \textit{n}-dimensional vector is represented as a $\mathbb{R^{N}}$ column matrix with all of its elements as real numbers ($\mathbb{R}$). All vector operations are then handled as matrix manipulations. Thus, a complex number is defined by the data-type $\mathbb{R}^{2}$, i.e., a column matrix having two elements. Similarly, a real number can be expressed as a 1-dimensional vector $\mathbb{R}^{1}$ or a number on a real line $\mathbb{R}$. In multivariate calculus theories of \holl, all theorems have been formally verified for functions with an arbitrary data-type $\mathbb{R^{N}} \to \mathbb{R^{M}}$. \\
We provide some of the frequently used \holl~functions in our proposed formalization.
\begin{definition}
\label{DEF:Cx}
{
\textup{\texttt{\textsf{
$\vdash_{def}$ $\forall$a. Cx a = complex (a,\&0)
}}}}
\end{definition}

The function \texttt{\textsf{Cx}} accepts a real number and returns its equivalent complex number with imaginary part equal to zero. Here the operator \texttt{\textsf{\&}} typecasts a natural number to its corresponding real number.

\begin{definition}
\label{DEF:Re_and_lift}
{
\textup{\texttt{\textsf{
$\vdash_{def}$ $\forall$z. Re z = z$\$$1  \\
$\mathtt{}$$\vdash_{def}$ $\forall$x. lift x = (lambda i. x)
}}}}
\end{definition}

The function \texttt{\textsf{Re}} accepts a complex number and returns its real part. The notation \texttt{\textsf{z$\$$i}} represents the $i^{th}$ component of a vector \texttt{z}. The function \texttt{\textsf{lift}} accepts a real number and maps it to a $1$-dimensional vector using \texttt{\textsf{lambda}} operator.

\begin{definition}
\label{DEF:Integral_Derivative}
{
\textup{\texttt{\textsf{
$\vdash_{def}$ $\forall$f i. integral i f = (@y.(f has\_integral y) i)   \\
$\mathtt{}$$\vdash_{def}$  $\forall$f net. vector\_derivative f net =  (@f'.(f has\_vector\_derivative f') net)
}}}}
\end{definition}

The function \texttt{\textsf{integral}} accepts an integrand function \texttt{\textsf{f:}}$\mathbb{R^{N}} \to \mathbb{R^{M}}$  and a vector space \texttt{\textsf{i:}}$\mathbb{R^{N}} \to \mathbb{B}$, which defines the region of integration and returns the corresponding vector integral. Here $\mathbb{B}$ represents the boolean data type. The function \texttt{\textsf{has\_integral}} defines the same relationship in the relational form.
The function \texttt{\textsf{vector\_derivative}} accepts a function \texttt{\textsf{f:}}$\mathbb{R}^{1} \to \mathbb{R^{M}}$ that needs to be differentiated and a \texttt{\textsf{net:}}$\mathbb{R}^{1} \to \mathbb{B}$ that defines the point at which the function \texttt{\textsf{f}} has to be differentiated and returns a vector of data-type $\mathbb{R^{M}}$ representing the differential of \texttt{\textsf{f}} at \texttt{\textsf{net}}.
The function \texttt{\textsf{has\_vector\_derivative}} defines the same relationship in the relational form. Here, the Hilbert choice operator \texttt{\textsf{@:}}$\mathbb{(\alpha \to \texttt{\textsf{bool}})\to  \alpha}$ returns values of the integral and differential, if they exist.


\subsection{Bond Graphs} \label{SUBSEC:bond_graphs}

\begin{figure}[!ht]
	\centering
	\resizebox{0.38\hsize}{!}{\includegraphics{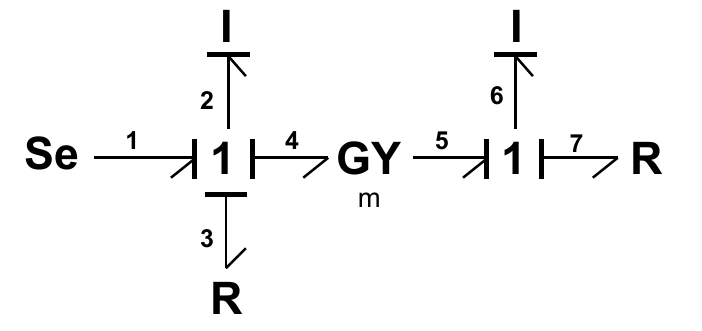}}
           \caption{Bond Graph Representation}
	\label{fig:bondgraph}
\end{figure}

BG~\cite{broenink1999introduction} is a directed graph composed of bonds and components that are connected together, as shown in Figure~\ref{fig:bondgraph}. The power bond is the most powerful component of a BG bridging any two components of BG and providing an exchanged power between them.

\begin{figure}[!ht]
	\centering
	\resizebox{0.35\hsize}{!}{\includegraphics{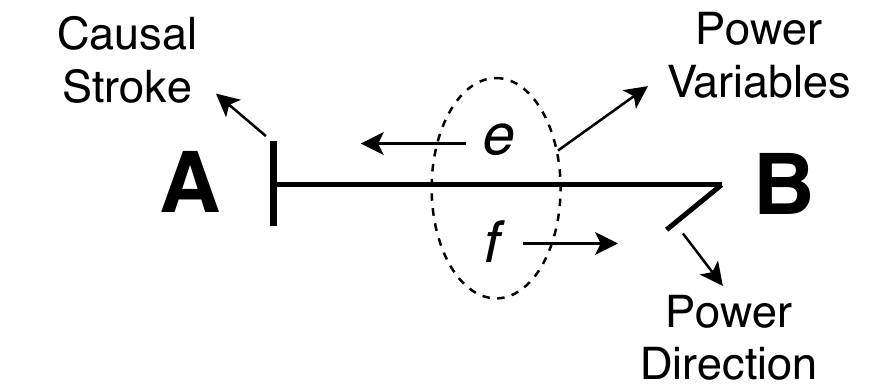}}
           \caption{Bond Representation}
	\label{fig:bond}
\end{figure}

A power bond is represented by a half arrow whose head indicates the direction of a positive power flow, as shown in Figure~\ref{fig:bond}. BG models are based on three types of analogies namely, signal, component and the connection analogies~\cite{gawthrop2007bond}.
An analogy is a mapping of the dynamical phenomenas/properties from one physical system to another. In BG, systems from different domains result into analogous equations, utilizing the concept of analogies.
\par
There are four types of signals in BG known as effort ($e$), flow ($f$), integrated effort ($p$) and integrated flow ($q$) signals.
These signals capture information about different aspects of systems from a wide range of domains/areas. For example, in the electrical domain, voltage ($e$), current ($f$), flux linkage ($p$) and charge ($q$) are represented by these signals~\cite{borutzky2011bond}, respectively. Since power is a product of effort and flow signals, a power bond is composed of effort and flow signals (variables), as shown in Figure~\ref{fig:bond}.

The other two signals, i.e., the integrated effort and flow signals belong to a class of energy variables. The integrated effort also known as generalized momentum, is mathematically expressed as a time integral over the effort signal.
\begin{equation} \label{p_eq}
p(t) = \int_{}^{t} e(t)dt = p_0 + \int_{0}^{t} e(t)dt
\end{equation}
\noindent Where $p_0$ represents the value of $p$ at time $t=0$. Similarly, the integrated flow also known as generalized displacement is mathematically described as a time integral of a flow signal.
\begin{equation} \label{q_eq}
q(t) = \int_{}^{t} f(t)dt = q_0 + \int_{0}^{t} f(t)dt
\end{equation}
\noindent Where $q_0$ represents the value of $p$ at time $t=0$. Therefore, the effort and flow signals are mathematically represented as the differentials of the generalized momentum and displacement, respectively.
\begin{equation} \label{e_eq}
e(t) = \dot p(t)
\end{equation}
\begin{equation} \label{f_eq}
f(t) = \dot q(t)
\end{equation}
The simulation of a BG is primarily based on the order of a computation of the effort and flow variables. This order is represented by a perpendicular bar (causal stroke) added to an end of a bond  (head or tail of the arrow, as shown in Figure~\ref{fig:bond}), thus indicating the corresponding variables acting as an input (cause) and an output (effect), respectively. This cause and effect phenomenon is generally known as causality, indicating the direction of the effort and flow signals (variables) in a BG model. In Figure~\ref{fig:bond}, effort $e$ is the input to A and the flow $f$ is the output from A. Similarly, $f$ acts as an input to B and $e$ acts as an output.

\begin{table}[!ht]
	\centering
           \resizebox{0.8\hsize}{0.6\textheight}{\includegraphics{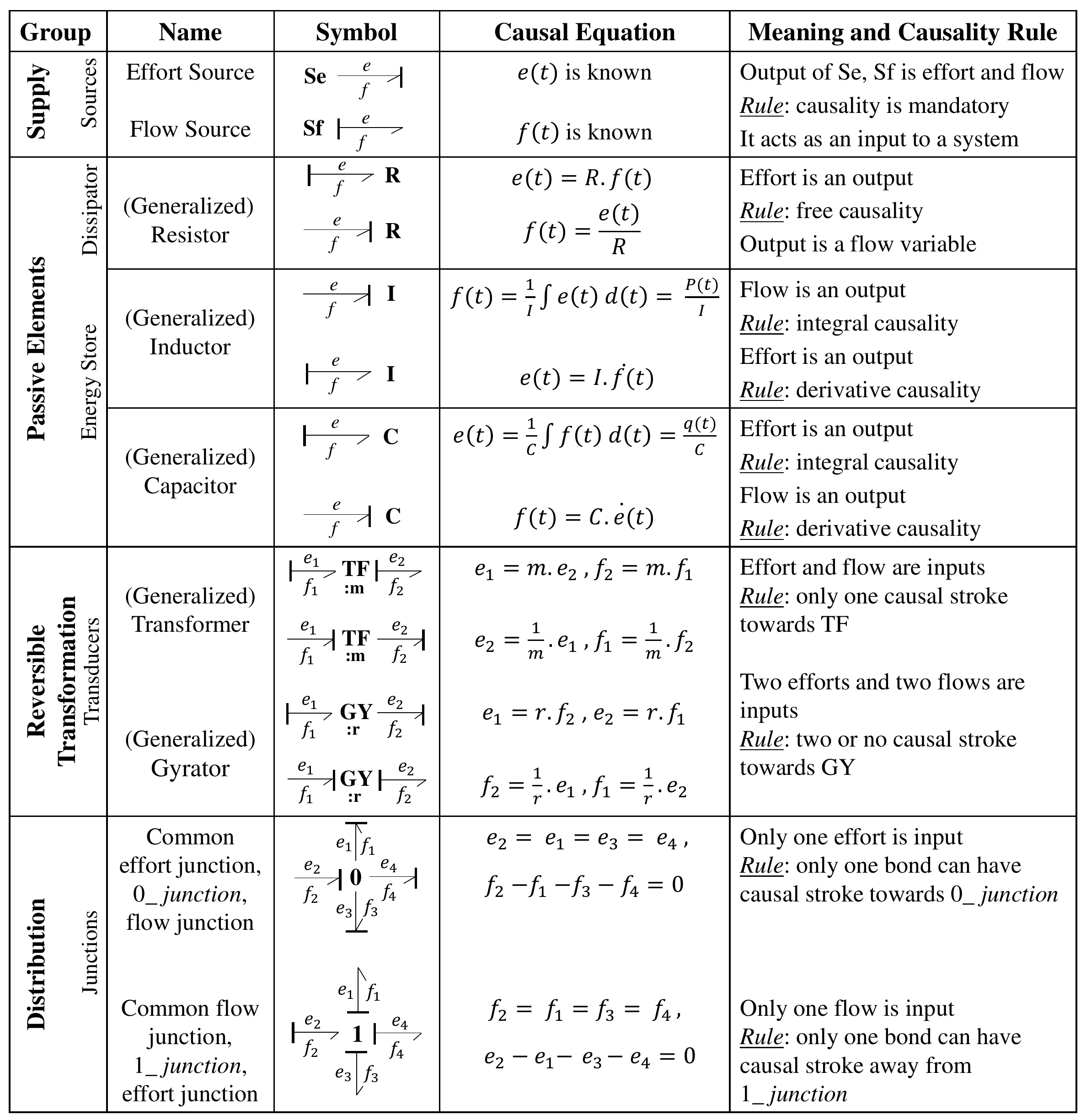}}
           \caption{Basic Components of Bond Graph}
	\label{tbl:basic_components_table}
\end{table}

A user needs to have some prior knowledge of causality, i.e., how to assign it manually, to construct the BG models of complex physical systems. Many approaches exist in the literature~\cite{broenink1999introduction,joseph1974method} for a systematic assignment of the casuality. However, Sequential Causality Assignment Procedure (SCAP) is generally preferred~\cite{borutzky2010derivation,broenink1999introduction} due to the systematic process and provides state-space representation of a system.
Causality analysis provides information about the various aspects of a system, i.e., inconsistency of design and an ill-posed model, the number of state-space variables and the presence of algebraic loops, which results in some complex Differential Algebraic Equations (DAEs), without deriving equations of a system.
\par
\begin{figure}[!ht]
 \begin{center}
 \begin{tabular}{cc}
  \resizebox{0.24\textwidth}{0.19\textheight}{\includegraphics{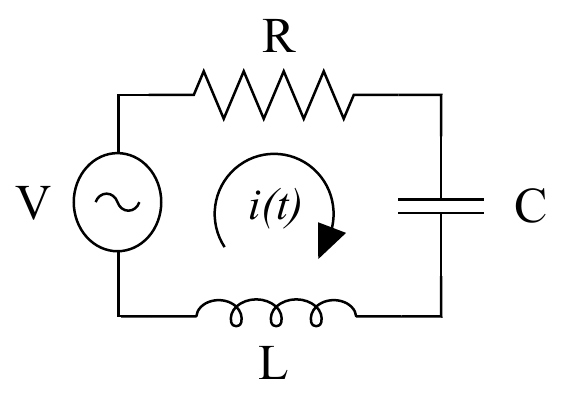}} &
  \resizebox{0.24\textwidth}{0.19\textheight}{\includegraphics{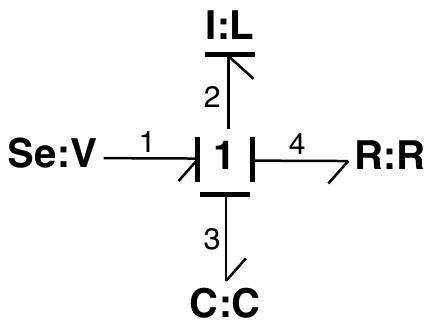}} \\
   {\small (a) RLC Circuit}  &
   {\small (b) Bond Graph}  \\
\end{tabular}
 \end{center}
\caption{RLC Circuit and its Corresponding Bond Graph Representation}
\label{fig:bg_exp}
\end{figure}

There are nine basic elements/components of a BG representation, categorized into four groups according to their characteristics, as shown in Table~\ref{tbl:basic_components_table}. Component analogies are further categorized into three groups, namely supply, passive elements and reversible transformation. The supply group contains sources of effort and flow variables. The passive elements contain the storage elements I and C as well as the dissipative element R. The reversible transformation group contains transducers TF and GY to convert one form of energy to another. Connection analogies consist of junctions $0$ and $1$ capturing the summation and equality laws. The first three groups of Table~\ref{tbl:basic_components_table} provide the component analogies and the last group represents the connection analogies.\\
Table~\ref{tab:defs} provides descriptions of various components of a BG representation, depicted in Figures~\ref{fig:bg_exp} and \ref{fig:novelbg}.
\newcolumntype{P}[1]{>{\centering\arraybackslash}p{#1}}
\newcolumntype{M}[1]{>{\centering\arraybackslash}m{#1}}
\begin{table}[!ht]
\footnotesize	
\renewcommand{\arraystretch}{1.00}
\begin{tabular}{|M{2.5cm} |M{13cm}|}
\hline
\textbf{Name} & \textbf{Description}\\
\hline
Strong Bond & A single bond that causes effort in the $0\_junction$ and flow in the $1\_junction$\\
\hline
Passive Element & A one port element that stores input power as potential energy (C-element), as kinetic energy (I-element) or transforms it into dissipative power (R-element)\\
\hline
Causal BG & A BG is called causally completed or causal if the causal stroke known as causality is added on one end of each bond\\
\hline
Causal Path & A sequence of bonds with/without a transformer in between having causality at the same end of all bonds or a sequence of bonds with a gyrator in between, and all the bonds of one side of the gyrator having same end causality while all the bonds on the other side with causality on opposite end. That means gyrator switches the direction of efforts/flows on one of its side~\cite{borutzky2010derivation}. A causal path can be a backward or forward or both depending upon the junction structure, elements and causality\\
\hline
Branch &  A branch is a series of junctions having parent-child relationship. Two different sequences of junctions can be connected with a common bond or two-port element. \mbox{Thus, one of the junction's sequence acts as parent branch and the other one as child}\\
\hline
Causal Loop & A causal loop is a closed causal path with bonds (of the child branch) either connected to a similar junction or two different junctions of the parent branch\\
\hline
\end{tabular}
\caption{Components of a Bond Graph Representation}
\label{tab:defs}
\end{table}
\noindent Now, we illustrate a BG based dynamic analysis of a most commonly used RLC circuit~\cite{kypuros2013system}, depicted in Figure~\ref{fig:bg_exp}a. To perform the BG based analysis of the RLC circuit, first we formally model the given circuit using its BG representation as shown in Figure~\ref{fig:bg_exp}b. Next, the laws of BGs, given in Table~\ref{tbl:basic_components_table}, are applied to obtain the corresponding set of equations. Finally, the BG components, given in Table~\ref{tab:defs}, are used to derive the  corresponding state-space model of the given RLC circuit. Generally, the BG representation capturing the dynamics of a system is based on transforming (mapping) system's components to their BG model counterparts and it varies according to the systems from various domains, such as, electrical, mechanical and medicine~\cite{thoma2016introduction}.\\
For the case of the given RLC circuit (Figure~\ref{fig:bg_exp}a), the voltage $V$ is modeled by the effort source $S_e$ and the inductor $L$ is modeled by the inductor $I$. Moreover, the capacitor ($C$) and the resistor ($R$) are mapped to the energy storing element C and passive element R, respectively, as shown in Figure~\ref{fig:bg_exp}b~\cite{broenink1999introduction}.

\noindent In the considered RLC circuit, same current flows from all components due to their series configuration. However, every component exhibits different voltages. The voltage and current are mapped to the effort and flow variables, respectively. The presence of $1\_junction$ shows that the value of flow variable (current) through all connected bonds is the same (by the Kirchhoff's Current Law or KCL~\cite{broenink1999introduction}), and the summation of effort variables (voltages) with power direction of bonds is equal to zero (by the Kirchhoff's Voltage Law or KVL~\cite{kypuros2013system}) as shown in Table~\ref{tbl:basic_components_table}.
\noindent Each bond is labelled by a number and a causality is assigned on every bond by following the SCAP approach. Thus, BG presented in Figure~\ref{fig:bg_exp}b is known as a causal BG. In $1\_junction$, only one bond is responsible for the flow variable (cause), known as strong bond, which is bond number 2 as illustrated in Figure~\ref{fig:bg_exp}b.
To obtain a mathematical model (state-space model) of the given BG representation, we need to apply laws of components and junctions given in Table~\ref{tbl:basic_components_table}. The mathematical equations of various components are expressed as follows:
\begin{equation}\label{src}
e_1(t) = V(t)
\end{equation}
\[f_2(t) = \frac{1}{L} \ast (p_0 + \int_{0}^{t} e_2(t)\ d(t))\]
By using Equation~(\ref{p_eq}), the inductor's equation can be written as:
\begin{equation}\label{ind}
f_2(t) = \frac{p_2(t)}{L}
\end{equation}
Similarly, an equation for the capacitor can be represented in terms of energy variable ($q$) by using Equation~(\ref{q_eq}).
\[e_3(t) = \frac{1}{C} \ast (q_0 +\int_{0}^{t} f_3(t) \ d(t))\]
\begin{equation}\label{cap}
e_3(t)= \frac {q_3(t)} {C}
\end{equation}
The output of the dissipative component ($R$) depends algebraically on the input as:
\begin{equation}\label{res}
e_4(t) = R \ . \ f_4(t)
\end{equation}
Now, we apply both laws of $1\_junction$, i.e., KCL and KVL for the case of RLC circuit. Since there is only one bond, i.e., Bond 1, which has a positive power direction, as shown in Figure~\ref{fig:bg_exp}b. Therefore, all of the effort variables except $e_1$ have a negative sign in the application of the summation law, i.e., KVL.
\begin{equation}\label{sum_law}
e_1 - e_2 - e_3 - e_4 = 0
\end{equation}
Similarly, since Bond 2 is a strong bond, as shown in Figure~\ref{fig:bg_exp}b, so the equality law, i.e., KCL for $1\_junction$, is mathematically expressed as follows:
\begin{equation}\label{equal_law}
f_2 = f_1 = f_3 = f_4
\end{equation}
Next, to obtain the state equations, i.e., equations of the energy storing components, we use equations of the components, i.e., Equations~(\ref{src}, \ref{cap}, \ref{res}) in Equation~(\ref{sum_law}).
\[V(t) - e_2(t) - \frac {1} {C}\ . \ q_3(t) -  R\ .\ f_4(t) = 0\]
The goal is to obtain the state equation for inductor (Bond 2) but all the entries of the above equation are in the form of generalized momentum ($p$) and generalized displacement ($q$) except variable $f_4(t)$. Thus, to convert $f_4(t)$ into an energy variable ($p$ or $q$), we follow the causal strokes of the bonds till we reach an energy storing ($I$, $C$) or a source ($S_e$, $S_f$) component with integral causality. In Figure~\ref{fig:bg_exp}b, we follow the causal strokes from Bond 4 to 2 and by back propagation $f_{2,4}$ of the causal strokes in junction structure ($1\_junction$), we can rewrite the above equation using Equation~\ref{equal_law} as follows:
\[V(t) - e_2(t) - \frac {1} {C}\ . \ q_3(t) -  R\ .\ f_2(t) = 0\]
By using Equations~\ref{e_eq} and \ref{ind}, the above equation can be rewritten as:
\begin{equation} \label{state_eq_1}
\dot p_2(t) = e_2(t) =  - \frac{R}{L}\ .\ p_2(t) - \frac {1} {C}\ .\ q_3(t) + V(t)
\end{equation}
Similarly, the equation for the storage component $C$ is as follows:
\[f_3(t) = f_2(t)\]
By using Equations~\ref{f_eq} and \ref{ind} in the above equation, we obtain the final form of the state equation for component C (Bond 3)  as follows:
\begin{equation} \label{state_eq_2}
\dot q_3(t) = f_3(t) = \frac{1}{L}\ .\ p_2(t)
\end{equation}
\noindent Finally, by applying various properties of vectors and matrices, Equations (\ref{state_eq_1}) and (\ref{state_eq_2}) can be transformed to their corresponding state-space models as follows:
\[\dot x(t) = A\ .\ x(t) + B\ .\ u(t)\]
\begin{equation} \label{state_mat}
\begin{bmatrix} \texttt{$\dot p_2(t)$} \\ \texttt{$\dot q_3(t)$}
\end{bmatrix} = \begin{bmatrix} \texttt{{\small $-\dfrac{R}{L} \ \ \ -\dfrac {1} {C}$}} \\
\texttt{{\small $\dfrac{1}{L} \ \ \ \ \ \ \ 0$}} \end{bmatrix} \begin{bmatrix} \texttt{$p_2(t)$} \\ \texttt{$q_3(t)$}
\end{bmatrix} + \begin{bmatrix} \texttt{$1$}\\ \texttt{$0$} \end{bmatrix} \begin{bmatrix} \texttt{$V(t)$} \end{bmatrix}
\end{equation}
The above state-space model corresponding to the BG representations is used for analyzing various properties of the underlying system, such as, stability.
\section{Proposed Methodology for the Bond Graph based Formal Analysis} \label{SEC:flow of bg}
\begin{wrapfigure}{r}{0.3\textwidth}
\graphicspath{ {./Images/} }
\includegraphics[width=1\linewidth]{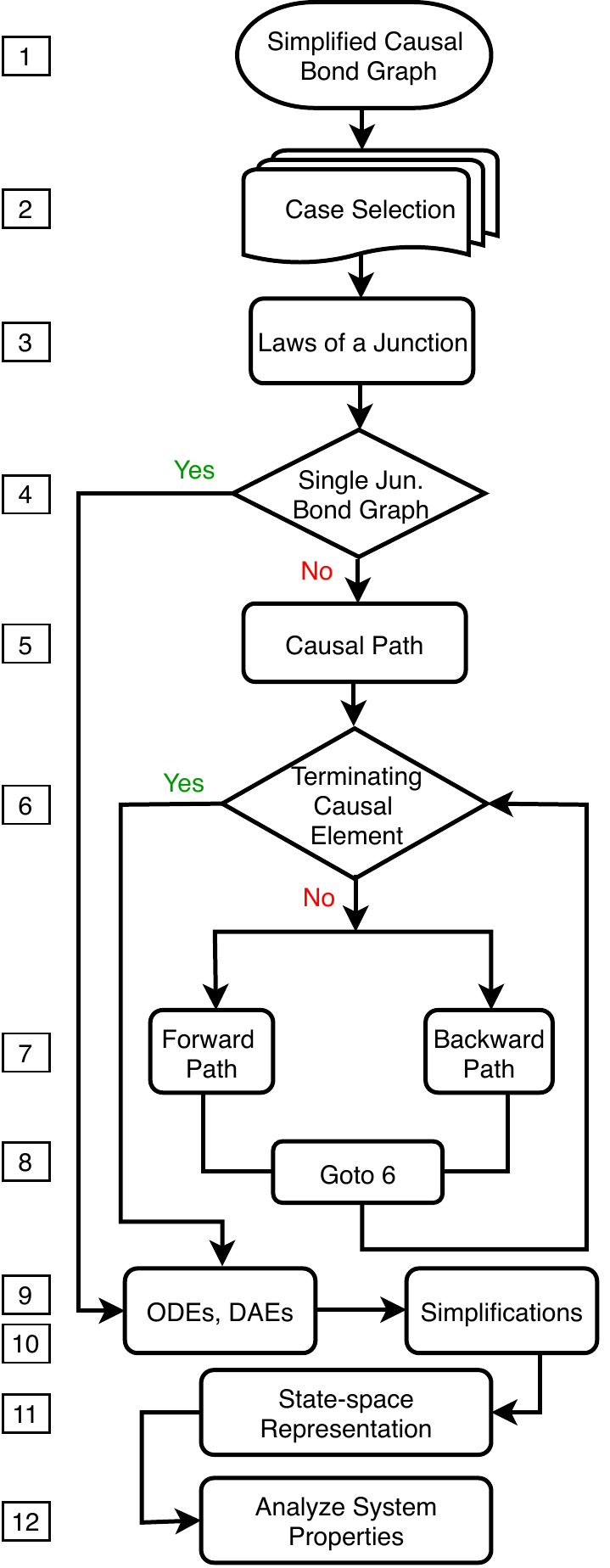}
\caption{Algorithm for Bond Graph based Analysis of Systems}
\label{fig:flowchart}
\end{wrapfigure}
This section provides the proposed methodology, depicted in Figure~\ref{fig:flowchart} for performing the BG based formal analysis of the dynamics of systems.\\
\textbf{Step 1:} Take a BG representation as an input from the user. The input BG should be:\\
(a) Causally completed\\ (b) Simplified after applying the simplification rules if necessary\\
\textbf{Step 2:} Apply different cases depending on the junction, its components and their causality.\\
\textbf{Step 3:} Apply laws of a junction, i.e., the equality or summation laws, one by one, based on the nature of the bond causality and junction.\\
\textbf{Step 4:} If the BG representation has only one junction, go to Step 9, otherwise move to the next step. \\
\textbf{Step 5:} Start following a causal path. \\
\textbf{Step 6:} When a terminating causal element is found within the same junction, go to Step 9, otherwise, go to the next step. \\
\textbf{Step 7:} Follow forward or backward path, or both, based on the causality of bonds and the nature of a junction. \\
\textbf{Step 8:} Repeat the above process by going back to Step 6.\\
\textbf{Step 9:} Extract Ordinary Differential Equations (ODEs) and Differential Algebraic Equations (DEAs) of the components of a junction.\\
\textbf{Step 10:} Simplify the differential equations to obtain the state equations.\\
\textbf{Step 11:} Generate state-space models from the state equations. \\
\textbf{Step 12:} Analyze various properties of the system, such as, stability.
\par
\begin{figure}[H]
 \begin{center}
 \begin{tabular}{c}
  \resizebox{0.52\textwidth}{!}{\includegraphics{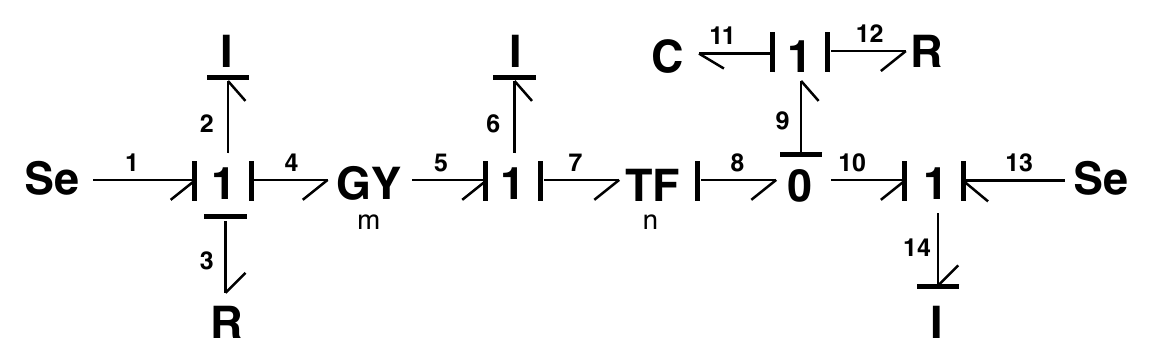}} \\
  {\small (a) Bond Graph Representation}\\ \\
  \resizebox{0.52\textwidth}{!}{\includegraphics{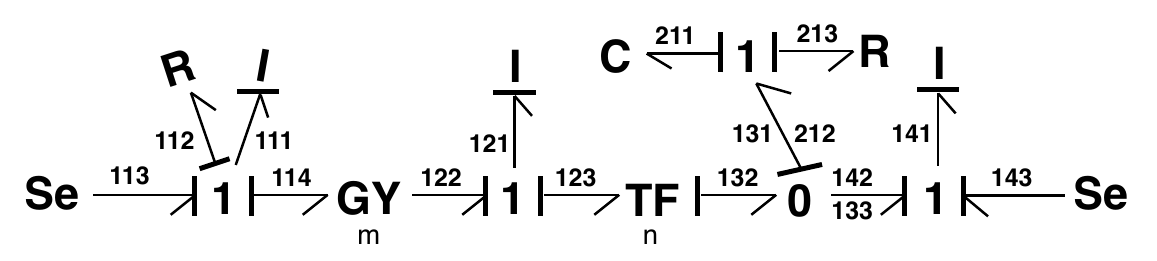}} \\
   {\small (b) Relabelling of Bond Graph Using Rules}  \\
\end{tabular}
 \end{center}
\caption{Representations of a Bond Graph}
\label{fig:novelbg}
\end{figure}

Now, we present the rules that are followed, to obtain a relabeled BG representation, during the formalization of the BGs in \holl. Figure~\ref{fig:novelbg} provides two different representations of an arbitrary BG, i.e., the given BG representation and the relabelled BG representation after applying the rules/assumptions to facilitate the corresponding formalization of the BGs. The rules for the formalization of the BG representation are as follows:\\
\textbf{1.} Arrange all one-port components/elements on the top of the corresponding junction and label all bonds of a junction in an anti-clockwise direction as shown in Figure~\ref{fig:novelbg}b. \\
\textbf{2.} A bond $\mathbf{B}_{pqr}$ is labelled according to its branch \texttt{p}, junction \texttt{q} and bond position \texttt{r} in a BG representation using integers. This applies to all bonds present in a BG representation.
For example, Bond 1 in Figure~\ref{fig:novelbg}a is relabelled as Bond 113, as shown in Figure~\ref{fig:novelbg}b. Bond 113 is the third bond of the first junction, which is an element of the first (parent) branch. Similarly, Bond 11 of Figure~\ref{fig:novelbg}a is relabelled as Bond 211 representing the first bond of the first junction that lies in the second (child) branch, as shown in Figure~\ref{fig:novelbg}b.
\noindent
\textbf{3.} One port (a connecting bond, Se/Sf, I, C, R) and two port (TF, GY) components/elements of a BG are also distinguished by using integers from 0 to 6, respectively. For example, in Figure~\ref{fig:novelbg}b, Bond 133 or 142 is a connecting bond/common bond (connecting 0 and 1 junction), so integer 0 is assigned to it in our formalization of BG. To recognize TF and GY, integer 5 and 6 are used respectively in \holl. Bond 113 and 143 are connected to a source component (Se), thus, the integer value 1 is assigned to it.  \\
\textbf{4.} The causality of a bond towards the respective junction is represented by the Boolean variables $\mathbf{T}$ and $\mathbf{F}$ for the causal stroke in the opposite direction of the junction. The causality of Bond 133 is $\mathbf{F}$ with respect to $0\_junction$, whereas, the causality of Bond 142 is $\mathbf{T}$ with respect to $1\_junction$. \\
\textbf{5.} The direction of a bond is a Boolean variable $\mathbf{T}$ when it is towards the junction and $\mathbf{F}$ for the opposite direction. Bond 113 has a positive direction, therefore, it is represented by $\mathbf{T}$ and Bond 112 is represented by $\mathbf{F}$ in \holl. \\
\textbf{6.} The modulus of a two-port component is represented in the form of a list containing the modulus of effort and flow variables. A two-port component (GY, TF) consists of two equations as described in Table~\ref{tbl:basic_components_table}. For example, in Figure~\ref{fig:novelbg}a, the list of the transformer component for the Bond 7 and 8 is in the form of \Big[$\dfrac{\texttt{1}}{\texttt{n}}, \texttt{n}$\Big] and \Big[\texttt{n}, $\dfrac{\texttt{1}}{\texttt{n}}$\Big], respectively.\\
\textbf{7.} $1\_junction$ is represented by a Boolean value $\mathbf{T}$ and $\mathbf{F}$ presents the $0\_junction$.\\
\textbf{8.} If a branch appears on a junction, its presence is represented by $\mathbf{T}$ and the absence by $\mathbf{F}$. Only $0\_junction$ has a branch so its presence in \holl~is represented by a Boolean variable $\mathbf{T}$. \\
\textbf{9.} If a junction has a single or multiple branches, the bond numbers associated with the connecting/common bonds of a parent branch are used to differentiate branches. Otherwise, integer 0 is used. The branch on $0\_junction$ is recognized by value 131, because Bond 131 is the connecting bond of both branches.
\section{Formalization of Bond Graph Representation} \label{SEC:Formalization of BG model}
This section provides formal definitions of the building blocks of BG that can be used to formalize a wide range of systems in higher-order logic. We have developed our formalization of BG using the foundational definitions presented in Section~\ref{SUBSEC:Mult_cal_theories}.
\subsection{Formal Model of a BG Representation} \label{sec:bg_rep}
We model a bond as a 6-tuple using the type abbreviation in \holl~as follows:
\begin{definition}
\label{DEF:Abbrevations}
{
\textup{\texttt{\textsf{
 new\_type\_abbrev (\texttt{\textsf{\textbf{"bond"}}},  \\
\hspace*{2.0cm} `:causality \# direction \# (branch \# num) \# ele\_type \# mod \# (effort \# flow)`) \\
 new\_type\_abbrev (\texttt{\textsf{\textbf{"bonds"}}}, `:(bond) list`)  \\
 new\_type\_abbrev (\texttt{\textsf{\textbf{"jun"}}}, `:jun\_num \# jun\_type \# bonds`)  \\
 new\_type\_abbrev (\texttt{\textsf{\textbf{"jun\_list"}}}, `:(jun) list`)
}}}}
\end{definition}

\noindent where the first element of the $6$-tuple captures the causality of the bond. Similarly, the second element, i.e., \texttt{\textsf{direction}} models power direction. The third element of the $6$-tuple, i.e., \texttt{\textsf{(branch \# num)}} provides the  information of a branch as a pair of Boolean and an integer data type, where the first element of pair represents branch presence and the second element models the position of common bond. The next two elements of the $6$-tuple, i.e., \texttt{\textsf{ele\_type}} and \texttt{\textsf{mod}} model the type of a component (it can take integer values from $0$ to $6$) and the modulus of the two-port elements in the form of a list consisting of values of the complex data type as described in Section~\ref{SEC:flow of bg}, respectively. The last element of the $6$-tuple is itself a pair of a function of type $\mathbb{R}^{1} \to \mathbb{R}^{2}$, providing the effort and flow of the bond. Similarly, the second type \texttt{\textsf{(bonds)}} provides a list of bonds of a single junction of a BG representation. The next type \texttt{\textsf{(jun)}} is a $3$-tuple composed of junction number (integer), type of a junction (it takes a Boolean value True and False for the junctions $1$ and $0$, respectively) and bonds of a junction. Finally, the type \texttt{\textsf{jun\_list}} provides a list of all the junctions that are present in a BG, thus capturing the overall BG representation of a system.


\subsection{Formalization of a Single Bond} \label{sec:bondrep}

As discussed in Section~\ref{SUBSEC:bond_graphs}, \textit{e} and \textit{f} are power variables while \texttt{\textsf{momentum}} and \texttt{\textsf{displacement}} represent energy variables. The formalized functions of these energy variables in \holl~are as follows:

\begin{definition}
\label{DEF:Basic_Concepts}
{
\textup{\texttt{\textsf{
$\vdash_{}$  $\forall$e f. power e f = ($\lambda$t. e t $\ast$ f t) \\
$\mathtt{}$$\vdash_{}$   $\forall$e. momentum e $\mathtt{p_0}$ = ($\lambda$t. $\mathtt{p_0}$ + integral (interval [lift(\&0), t]) e) \\
$\mathtt{}$$\vdash_{}$  $\forall$f. displacement f $\mathtt{q_0}$ = ($\lambda$t. $\mathtt{q_0}$ + integral (interval [lift(\&0), t]) f)  \\
$\mathtt{}$$\vdash_{}$   $\forall$e. momentum\_der e $\mathtt{p_0}$ = ($\lambda$t. vector\_derivative (momentum e $\mathtt{p_0}$) (at t)) \\
$\mathtt{}$$\vdash_{}$ $\forall$f. displacement\_der f $\mathtt{q_0}$ = ($\lambda$t. vector\_derivative (displacement f $\mathtt{q_0}$) (at t))
}}}}
\end{definition}

Now, we formalize the causality of a bond in \holl~as the following recursive functions.

\begin{definition}
\label{DEF:causality_dir}
{
\textup{\texttt{\textsf{
$\vdash_{}$  $\forall$j i k. ($\mathtt{\mathsf{causality_f}}$ j i 0 = 0) $\land$  ($\mathtt{\mathsf{causality_f}}$ j i (SUC k) =  \\
\hspace*{4.3cm}  (if causality j i (SUC k) = F then SUC k else $\mathtt{\mathsf{causality_f}}$ j i k))  \\
$\vdash_{}$  $\forall$j i k. ($\mathtt{\mathsf{causality_t}}$ j i 0 = 0) $\land$  ($\mathtt{\mathsf{causality_t}}$ j i (SUC k) =   \\
\hspace*{4.3cm}  (if causality j i (SUC k) = T then SUC k else $\mathtt{\mathsf{causality_t}}$ j i k))
}}}}
\end{definition}

The $\mathtt{\mathsf{causality_f}}$ accept a  list of junction \texttt{\textsf{j}}, an element \texttt{\textsf{i}} of the list \texttt{\textsf{j}} and a bond \texttt{\textsf{k}} of the junction \texttt{\textsf{i}}, and returns the causality of the bond \texttt{\textsf{k}}, i.e., it returns a Boolean value \texttt{\textsf{T}} for the case of direction of bond away from a junction, as described in Section~\ref{SEC:flow of bg}. Similarly, the function $\mathtt{\mathsf{causality_t}}$ returns true (\texttt{\textsf{T}}) if the direction of a bond \texttt{\textsf{k}} is towards a junction \texttt{\textsf{i}}. \\
A power bond carries information of causality as well as its power direction as shown in Figure~\ref{fig:bond}. Its formalized function is as follows:

\begin{definition}
\label{DEF:bond_dir}
{
\textup{\texttt{\textsf{
$\vdash_{}$  $\forall$j i k. bond\_direction\_cond j i k =   \\
\hspace*{4.5cm}  (if bond\_direction j i k = T then Cx (\&1) else  -- Cx (\&1))
}}}}
\end{definition}

The function \texttt{\textsf{bond\_direction}} accepts a list of junctions \texttt{\textsf{j}}, an element \texttt{\textsf{i}} of \texttt{\textsf{j}} and a bond \texttt{\textsf{k}} of the junction \texttt{\textsf{i}} and returns \texttt{\textsf{T}} for a positive power direction, otherwise it returns false. It basically extracts the second element from the $6$-tuple capturing a BG representation. Moreover, the function \texttt{\textsf{bond\_direction\_cond}} assigns the complex numbers $1$ and $-1$ to the Boolean values \texttt{\textsf{T}} and \texttt{\textsf{F}}, respectively.


\subsection{Formalization of Basic Components of a Single Bond and its Associated Laws} \label{sec:components}

In BGs, sources are called active elements with one variable equal to zero. $\mathtt{\mathsf{src_e}}$ represents supply effort to the system and $\mathtt{\mathsf{src_f}}$ represents supply flow to the system and are defined as follows:

\begin{definition}
\label{DEF:sources}
{
\textup{\texttt{\textsf{
$\vdash_{}$  $\forall$e. $\mathtt{src_e}$ e = ($\lambda$t. Cx e)  \\
$\mathtt{}$$\vdash_{}$  $\forall$f. $\mathtt{src_f}$ f = ($\lambda$t. Cx f)
}}}}
\end{definition}

These are the passive elements of BG as described in Table~\ref{tbl:basic_components_table}. Both integral and differential causal elements are formalized. $\mathtt{\mathsf{compliance_e}}$ represents integrally causal C-element and $\mathtt{\mathsf{compliance_f}}$ has differential causality.

\begin{definition}
\label{DEF:assive_ele}
{
\textup{\texttt{\textsf{
$\vdash_{}$  $\forall$R f. $\mathtt{res_e}$ R f = ($\lambda$t. Cx R $\ast$ f t)   \\
$\mathtt{}$$\vdash_{}$  $\forall$R e. $\mathtt{res_f}$ R e = ($\lambda$t. Cx  \Big($\dfrac{\texttt{\&1}}{\texttt{R}}\Big) \ \ast$ e t)  \\
$\vdash_{}$  $\forall$C f. $\mathtt{compliance_e}$ C f $\mathtt{q_0}$ = ($\lambda$t. Cx \Big($\dfrac{\texttt{\&1}}{\texttt{C}}\Big) \ \ast$ displacement f $\mathtt{q_0}$ t) \\
$\mathtt{}$$\vdash_{}$  $\forall$C e. $\mathtt{compliance_f}$ C e = ($\lambda$t. Cx C $\ast$ vector\_derivative e (at t))  \\
$\vdash_{}$  $\forall$L f. $\mathtt{inductor_e}$ L f = ($\lambda$t. Cx L $\ast$ vector\_derivative f (at t))   \\
$\mathtt{}$$\vdash_{}$  $\forall$L e. $\mathtt{inductor_f}$ L e $\mathtt{p_0}$ = ($\lambda$t. Cx \Big($\dfrac{\texttt{\&1}}{\texttt{L}}\Big)\ \ast$ momentum e $\mathtt{p_0}$ t)
}}}}
\end{definition}

Next, we provide the equality laws for the junctions $0$ and $1$  (presented as Equation (\ref{equal_law}) for RLC circuit example in Section~\ref{SUBSEC:bond_graphs}) of a BG representation in \holl~as follows:

\begin{definition}
\label{DEF:equality_laws}
{
\textup{\texttt{\textsf{
$\vdash_{}$  $\forall$j i k1 k2 t. jun\_0\_e j i k1 k2 t =   \\
\hspace*{6.5cm}  (bond\_effort j i k1 t = bond\_effort j i k2 t)   \\
$\mathtt{}$$\vdash_{}$  $\forall$j i k1 k2 t. jun\_1\_f j i k1 k2 t = (bond\_flow j i k1 t = bond\_flow j i k2 t)
}}}}
\end{definition}

The functions \texttt{\textsf{jun\_0\_e}} and \texttt{\textsf{jun\_1\_f}} represent laws of equality for junctions $0$ and $1$, respectively. The function \texttt{\textsf{jun\_0\_e}} accepts a list of junctions \texttt{\textsf{j:(jun) list}}, an element \texttt{\textsf{i}} of the list \texttt{\textsf{j}} representing a junction ($0\_junction$), two different bonds \texttt{\textsf{k1}} and \texttt{\textsf{k2}} of the junction \texttt{\textsf{i}}, and models law of equality as an equation of effort. It uses the function \texttt{\textsf{bond\_effort}} to capture the effort (e) variable. Similarly, the function \texttt{\textsf{jun\_1\_f}} models law of equality as an equation of flow for $1\_junction$ using the function \texttt{\textsf{bond\_flow}}.
\par
Next, we formalize the summation laws for the junctions $0$ and $1$ of a BG representation in \holl~as follows:

\begin{definition}
\label{DEF:summation_laws}
{
\textup{\texttt{\textsf{
$\vdash_{}$  $\forall$j i t. jun\_1\_e j i t = vsum (0$\ldots$bonds\_length j i - 3)   \\
\hspace*{10cm}  ($\lambda$k. bond\_effort\_wd j i k t)    \\
$\mathtt{}$$\vdash_{}$  $\forall$j i t. jun\_0\_f j i t = vsum (0$\ldots$bonds\_length j i - 3) ($\lambda$k. bond\_flow\_wd j i k t)
}}}}
\end{definition}

The function \texttt{\textsf{jun\_1\_e}} accepts a list of junctions \texttt{\textsf{j:(jun) list}}, an element \texttt{\textsf{i}} of the list \texttt{\textsf{j}}, representing a $1\_junction$, and a time variable \texttt{\textsf{t}} and returns the summation of the effort variables of all bonds in the junction i, i.e., $1\_junction$ except the last two bonds. Here, the function \texttt{\textsf{bond\_effort\_wd}} models the effort variable of a bond with power direction. Similarly, the function \texttt{\textsf{jun\_0\_f}} represents summation law of $0\_junction$ by skipping the last two bonds. Moreover, any two junctions are connected by a common bond, For example, junctions 0 and 1 are connected by Bond 10 as shown in Figure~\ref{fig:novelbg}a by the laws of summation and equality. Their flow variables for both junctions are mathematically expressed as:
\[f_8(t) - f_9(t) - f_{10}(t) = 0, \ \  f_{8}(t) = n \ast f_{7}(t), \ \ f_{10}(t) = f_{14}(t)\]
We can simplify above equations by substituting the value of flow variables of common bond and transformer (TF), i.e., Bonds $8$ and $10$, from second and third equations to first equation, resulting into the following equation:
\[n \ast f_7(t) - f_9(t) - f_{14}(t) = 0\]
So, in order to obtain the final form of the equations, we eliminate these intermediate steps (substitution of equations) by excluding last two bonds of a junction.

\begin{definition}
\label{DEF:modulus_cond}
{
\textup{\texttt{\textsf{
$\vdash_{}$  $\forall$j i k. modulus\_cond j i k = (if causality j i k = T   \\
\hspace*{5.0cm}  then EL 1 (modulus j i k) else EL 0 (modulus j i k))
}}}}
\end{definition}

A two-port element consists of two equations with different modulus for different variables (effort and flow), as shown in Table~\ref{tbl:basic_components_table}. The function \texttt{\textsf{modulus\_cond}} verifies the causality of a bond towards a junction and provides the first element (flow variable) of the list \texttt{\textsf{modulus}} accordingly. Similarly, if the causality of a bond is not towards the junction then the modulus of the effort variable is returned.


\subsection{Formalization of the Components of a BG} \label{sec:bgcomponents}

We formalize the strong bond in \holl~as follows:

\begin{definition}
\label{DEF:strong_bond}
{
\textup{\texttt{\textsf{
$\vdash_{}$   $\forall$j i. strong\_bond\_f j i = $\mathtt{\mathsf{causality_f}}$ j i (bonds\_length j i - 1) \\
$\vdash_{}$  $\forall$j i. strong\_bond\_t j i = $\mathtt{\mathsf{causality_t}}$ j i (bonds\_length j i - 1)
}}}}
\end{definition}

The functions \texttt{\textsf{strong\_bond\_f}} and \texttt{\textsf{strong\_bond\_t}} use the functions of Definition \ref{DEF:causality_dir}~namely $\mathtt{\mathsf{causality_f}}$ and $\mathtt{\mathsf{causality_t}}$ that check all bonds of a junction to detect the strong bonds with casualty towards a junction and in the opposite direction of a junction, respectively. Here, the function \texttt{\textsf{bonds\_length}} provides the total number of bonds of a single junction.
\par
Next, we model the two skipped bonds of the last junction as follows:

\begin{definition}
\label{DEF:last_jun_bond}
{
\textup{\texttt{\textsf{
$\vdash_{}$  $\forall$j i t. bw\_last\_jun\_e j i t = (if i = 0  \\
\hspace*{4.2cm} then snd\_last\_bond\_dir j i $\ast$ snd\_last\_bond\_e j i t else Cx (\&0))   \\
$\vdash_{}$  $\forall$j i t. bw\_last\_jun\_f j i t = (if i = 0  \\
\hspace*{4.2cm}  then snd\_last\_bond\_dir j i $\ast$ snd\_last\_bond\_f j i t else Cx (\&0))  \\
$\vdash_{}$  $\forall$j i t. fw\_last\_jun\_e j i t = (if i = 0  \\
\hspace*{4.2cm}  then last\_bond\_dir j i $\ast$ last\_bond\_e j i t else Cx (\&0))  \\
$\vdash_{}$  $\forall$j i t. fw\_last\_jun\_f j i t = (if i = 0   \\
\hspace*{4.2cm}  then last\_bond\_dir j i $\ast$ last\_bond\_f j i t else Cx (\&0))
}}}}
\end{definition}

In the summation laws, provided as Definition~\ref{DEF:summation_laws}, we skipped the last two bonds of a junction that need to be modeled. By following a causal path (backward and forward), if we reach at the first junction (in backward path) or the last junction (in forward path), i.e., $i = 0$ of a list \texttt{\textsf{j}}, we must add those skipped bonds, i.e., the second last and last bond, of junction \texttt{\textsf{i}} with a power direction, otherwise we need to add a zero. The function \texttt{\textsf{bw\_last\_jun\_e}} provides the effort variable with power direction modeled as \texttt{\textsf{snd\_last\_bond\_dir}} of the second last bond of the junction \texttt{\textsf{i}}, when following a backward causal path. Similarly, following a forward causal path, the function \texttt{\textsf{fw\_last\_jun\_e}} provides the effort variable with power direction modeled as \texttt{\textsf{last\_bond\_dir}} of the last bond of the junction \texttt{\textsf{i}}.

\begin{definition}
\label{DEF:last_jun_bonds_sum}
{
\textup{\texttt{\textsf{
$\vdash_{}$  $\forall$j i t. bw\_jun\_e\_sum j i t = (jun\_1\_e j i t) + (bw\_last\_jun\_e j i t)   \\
$\vdash_{}$  $\forall$j i t. bw\_jun\_f\_sum j i t = (jun\_0\_f j i t) + (bw\_last\_jun\_f j i t)  \\
$\vdash_{}$  $\forall$j i t. fw\_jun\_e\_sum j i t = (jun\_1\_e j i t) + (fw\_last\_jun\_e j i t)  \\
$\vdash_{}$  $\forall$j i t. fw\_jun\_f\_sum j i t = (jun\_0\_f j i t) + (fw\_last\_jun\_f j i t)
}}}}
\end{definition}

The functions \texttt{\textsf{bw\_jun\_e\_sum}} and \texttt{\textsf{fw\_jun\_e\_sum}} accept a list of junctions \texttt{\textsf{j:(jun) list}}, an element \texttt{\textsf{i}} of the list \texttt{\textsf{j}} and a time variable \texttt{\textsf{t}} and returns the summation of the effort variables of $1\_junction$ (last junction in the junction list \texttt{\textsf{j}}) including the second last and last bonds of the junction \texttt{\textsf{i}}, when following a backward and forward paths, respectively. Similarly, the functions \texttt{\textsf{bw\_jun\_f\_sum}} and \texttt{\textsf{fw\_jun\_f\_sum}} provide summation of the flow variables of $0\_junction$ including the second last and last bonds of the junction \texttt{\textsf{i}}, following the backward and forward paths.

\begin{definition}
\label{DEF:bw_path}
{
\textup{\texttt{\textsf{
$\vdash_{}$  $\forall$j i t. (backwrd\_path j 0 t = Cx(\&0)) $\land$   \\
\hspace*{1.0cm}   (backwrd\_path j (SUC i) t = if P1 then P2 else if P3 then P4 else Cx (\&0))
}}}}
\end{definition}

\noindent To find a causal path, we need to know the starting and the upcoming junction of the path, type of the common bond between two junctions and a strong bond. The function \texttt{\textsf{backward\_path}} is made up of four different parts whose explanation is provided in the Table~\ref{tab:bw_path}.\\
The first part, i.e., \texttt{\textsf{P1}} of the function \texttt{\textsf{backward\_path}} specifies a causal path starting from an arbitrary junction \texttt{\textsf{i}} (of any type, i.e., $0\_junction$ or $1\_junction$)  and a $0\_junction$ (modeled using Boolean value \texttt{\textsf{F}}) as an upcoming junction \texttt{\textsf{i}} in the path and both these junctions (junction \texttt{\textsf{i}} and $0\_junction$) share a common bond or a two-port element. The function \texttt{\textsf{snd\_last\_bond\_type}} captures the type of sharing bond. In \texttt{\textsf{P2}}, the function \texttt{\textsf{strong\_bond\_t}} checks if the strong bond, with causality towards the junction, of the upcoming junction \texttt{\textsf{i}} is the last, second last or an arbitrary bond alongwith its position in the list of junction \texttt{\textsf{j}}.
If the strong bond with causality towards the junction \texttt{\textsf{i}} (\texttt{\textsf{strong\_bond\_t}}) is the last bond of the junction \texttt{\textsf{i}} and the value of \texttt{\textsf{i}} is not equal to zero, i.e., it is not the first junction, the function \texttt{\textsf{bw\_jun\_f\_sum}} applies the summation law on the junction \texttt{\textsf{i}} alongwith power direction \texttt{\textsf{last\_bond\_dir\_cond}} and the modulus of last bond \texttt{\textsf{last\_bond\_modulus\_cond}}, and keeps following the \texttt{\textsf{backwrd\_path}}, until we reach a terminating element. The part \texttt{\textsf{P3}} of the function \texttt{\textsf{backward\_path}} is quite similar to part \texttt{\textsf{P1}} of the function, except the upcoming junction is now $1\_junction$. Finally, part \texttt{\textsf{P4}} behaves in just like part \texttt{\textsf{P2}}, using \texttt{\textsf{strong\_bond\_f}} for the causality of a strong bond in the opposite direction of the junction \texttt{\textsf{i}}. In other words, the function \texttt{\textsf{backwrd\_path}} consists of different combinations of junctions ($0$ or $1$), common bonds (connecting bond, TF or GY) and strong bonds (with causality towards or in the opposite direction of a junction), and provides their equations accordingly. The detailed definition can be found in our \holl~proof script~\cite{formalizationbg}.

\newcolumntype{P}[1]{>{\centering\arraybackslash}p{#1}}
\newcolumntype{M}[1]{>{\centering\arraybackslash}m{#1}}
\begin{table}[!ht]
\centering
\footnotesize
\renewcommand{\arraystretch}{1.00}
\begin{tabular}{|M{0.6cm} |M{14.8cm}|}
\hline
\textbf{Part} & \textbf{Formalized Form}\\
\hline
\texttt{\textsf{P1}} & {\raggedright \texttt{\textsf{(type\_of\_jun j (SUC i) = F) $\land$ (snd\_last\_bond\_type j (SUC i) = 0 $\lor$}} \\
\hspace*{1.35cm} \texttt{\textsf{snd\_last\_bond\_type j (SUC i) = 5) $\land$ (type\_of\_jun j i = F) $\lor$}} \\
\hspace*{0.2cm} \texttt{\textsf{(type\_of\_jun j (SUC i) = F) $\land$}} \texttt{\textsf{(snd\_last\_bond\_type j (SUC i) = 6) $\land$ (type\_of\_jun j i) = F)}}} \\
\hline
\texttt{\textsf{P2}} & {\raggedright \texttt{\textsf{if (strong\_bond\_t j i = snd\_last\_bond\_of\_jun j i) $\land \ \sim$(i = 0)}} \\
\hspace*{0.3cm} \texttt{\textsf{then last\_bond\_modulus\_cond j i $\ast$ backwrd\_path j i t}} \\
\hspace*{0.3cm} \texttt{\textsf{else if (strong\_bond\_t j i = last\_bond\_of\_jun j i) $\land \ \sim$(i = 0)}} \\
\hspace*{1.0cm} \texttt{\textsf{then last\_bond\_dir\_cond j i $\ast$ last\_bond\_modulus\_cond j i $\ast$ }} \\
\hspace*{3.43cm} \texttt{\textsf{(bw\_jun\_f\_sum j i t + backwrd\_path j i t)}} \\
\hspace*{1.0cm} \texttt{\textsf{else last\_bond\_modulus\_cond j i $\ast$}}  \texttt{\textsf{bond\_effort j i (strong\_bond\_t j i) t}}} \\
\hline
\texttt{\textsf{P3}} & {\raggedright  \texttt{\textsf{(type\_of\_jun j (SUC i) = T) $\land$ (snd\_last\_bond\_type j (SUC i) = 6) $\land$}} \\
\hspace*{1.9cm} \texttt{\textsf{(type\_of\_jun j i = T) $\lor$}}  \texttt{\textsf{(type\_of\_jun j (SUC i) = F) $\land$ }}   \\
\hspace*{1.5cm} \texttt{\textsf{(snd\_last\_bond\_type j (SUC i) = 6) $\land$ (type\_of\_jun j i) = T)}}} \\
\hline
\texttt{\textsf{P4}}  & {\raggedright  \texttt{\textsf{if (strong\_bond\_f j i = snd\_last\_bond\_of\_jun j i) $\land \ \sim$(i = 0)}}  \\
\hspace*{0.3cm} \texttt{\textsf{then last\_bond\_modulus\_cond j i $\ast$ backwrd\_path j i t}} \\
\hspace*{0.3cm} \texttt{\textsf{else if (strong\_bond\_f j i = last\_bond\_of\_jun j i) $\land \ \sim$(i = 0)}} \\
\hspace*{1.0cm} \texttt{\textsf{then last\_bond\_dir\_cond j i $\ast$ last\_bond\_modulus\_cond j i $\ast$}} \\
\hspace*{3.43cm} \texttt{\textsf{(bw\_jun\_e\_sum j i t + backwrd\_path j i t)}} \\
\hspace*{1.0cm} \texttt{\textsf{else last\_bond\_modulus\_cond j i $\ast$ bond\_flow j i (strong\_bond\_f j i) t)}}} \\
\hline
\end{tabular}
\caption{Formalized Parts of Definition~\ref{DEF:bw_path}}
\label{tab:bw_path}
\end{table}

\begin{definition}
\label{DEF:fw_path}
{
\textup{\texttt{\textsf{
$\vdash_{}$  $\forall$j i t. (fwrd\_path (j:jun\_list) 0 t = Cx(\&0)) $\land$  \\
\hspace*{1.0cm}  (fwrd\_path j (SUC i) t = if P1 then P2 else if P3 then P4 else Cx (\&0))
}}}}
\end{definition}

\noindent The function \texttt{\textsf{fwrd\_path}} accepts a list of junctions \texttt{\textsf{j}}, an element \texttt{\textsf{i}} of the list \texttt{\textsf{j}} and a time variable \texttt{\textsf{t}} and returns variable (effort or flow) or summation equation depending upon the structure of junctions, causality and elements. \\
The parts \texttt{\textsf{P1}} and \texttt{\textsf{P3}} of the function \texttt{\textsf{fwrd\_path}} are quite similar to that of \texttt{\textsf{backwrd\_path}} except the reversed list of junctions \texttt{\textsf{(rev j)}}. Moreover, parts \texttt{\textsf{P2}} and \texttt{\textsf{P4}} are quite similar to that of \texttt{\textsf{backwrd\_path}} except the placement of last and second last bonds. The detailed formalized form of the parts \texttt{\textsf{P1, P2, P3}} and \texttt{\textsf{P4}} is given in the Table~\ref{tab:fw_path}.

\begin{definition}
\label{DEF:match_jun}
{
\textup{\texttt{\textsf{
$\vdash_{}$  $\forall$j n i. jun\_num\_match j 0 i = jun\_num (rev j) 0 $\land$ \\
\hspace*{1cm}  jun\_num\_match j (SUC n) i =  \\
\hspace*{1.2cm} (if i = jun\_num (rev j) (SUC n) then SUC n else  jun\_num\_match j n i) \\
$\vdash_{}$  $\forall$j i. jun\_match j i = jun\_num\_match j (LENGTH j - 1) i  \\
$\vdash_{}$  $\forall$j i t. final\_fwrd\_path j i t = fwrd\_path j (jun\_match j i) t
}}}}
\end{definition}

Since, the function \texttt{\textsf{final\_fwrd\_path}} reverses the list of junctions \texttt{j}, thus the position of each entry (junction) of the list \texttt{j} changes in this process. However, we placed junction number \texttt{jun\_num} in the 6-tuple as described in Definition~\ref{DEF:Abbrevations}, which remains unchanged in the reversion process. Thus for accuracy, it is necessary to match the $i^{th}$ junction of the reversed list with the \texttt{jun\_num}
by using the function \texttt{jun\_num\_match}. There is only one junction in the reversed list that has a matching junction number, as each junction of a BG representation is assigned a unique number.
The function \texttt{jun\_match} checks all the junctions of list \texttt{j} for matching a junction and the final definition of forward path \texttt{final\_fwrd\_path} uses that resulting junction.
\newcolumntype{P}[1]{>{\centering\arraybackslash}p{#1}}
\newcolumntype{M}[1]{>{\centering\arraybackslash}m{#1}}
\begin{table}[!ht]
\centering
\footnotesize
\renewcommand{\arraystretch}{1.00}
\begin{tabular}{|M{0.6cm} |M{14.8cm}|}
\hline
\textbf{Part} & \textbf{Formalized Form}\\
\hline
\texttt{\textsf{P1}} & {\raggedright   \texttt{\textsf{(type\_of\_jun (rev j) (SUC i) = F) $\land$}} \\
\hspace*{0.8cm} \texttt{\textsf{(last\_bond\_type (rev j) (SUC i) = 0 $\lor$}}
   \texttt{\textsf{last\_bond\_type (rev j) (SUC i) = 5) $\land$ \\
\hspace*{0.5cm}  (type\_of\_jun (rev j) i = F) $\lor$}}
   \texttt{\textsf{(type\_of\_jun (rev j) (SUC i) = F) $\land$}}  \\
\hspace*{0.5cm}\texttt{\textsf{(last\_bond\_type (rev j) (SUC i) = 6) $\land$ (type\_of\_jun (rev j) i = F)}}} \\
\hline
\texttt{P2} & {\raggedright   \texttt{\textsf{if (strong\_bond\_t (rev j) i = last\_bond\_of\_jun (rev j) i) $\land \ \sim$(i = 0)}} \\
\hspace*{0.3cm} \texttt{\textsf{then snd\_last\_bond\_modulus\_cond (rev j) i $\ast$ fwrd\_path j i t}} \\
\hspace*{0.3cm} \texttt{\textsf{else if (strong\_bond\_t (rev j) i =}} \texttt{\textsf{snd\_last\_bond\_of\_jun (rev j) i) $\land \ \sim$(i = 0)}}  \\
\hspace*{0.5cm} \texttt{\textsf{then snd\_last\_bond\_dir\_cond j i $\ast$ }}  \\
\hspace*{1.0cm} \texttt{\textsf{snd\_last\_bond\_modulus\_cond (rev j) i $\ast$}} \texttt{\textsf{(fw\_jun\_f\_sum (rev j) i t + fwrd\_path j i t)}} \\
\hspace*{0.5cm} \texttt{\textsf{else snd\_last\_bond\_modulus\_cond (rev j) i $\ast$ }} \texttt{\textsf{bond\_effort (rev j) i (strong\_bond\_t (rev j) i) t}}} \\
\hline
\texttt{P3} & {\raggedright  \texttt{\textsf{(type\_of\_jun (rev j) (SUC i) = T) $\land$ (last\_bond\_type (rev j) (SUC i) = 6) $\land$}}  \\
\texttt{\textsf{(type\_of\_jun (rev j) i) = T) $\lor$}} \texttt{\textsf{(type\_of\_jun (rev j) (SUC i) = F) $\land$}}  \\
\texttt{\textsf{(last\_bond\_type (rev j) (SUC i) = 6) $\land$ (type\_of\_jun (rev j) i) = T)}}} \\
\hline
\texttt{P4} & {\raggedright   \texttt{\textsf{if (strong\_bond\_f (rev j) i = last\_bond\_of\_jun (rev j) i) $\land \ \sim$(i = 0)}} \\
\hspace*{0.3cm} \texttt{\textsf{then snd\_last\_bond\_modulus\_cond (rev j) i) $\ast$ fwrd\_path j i t}}  \\
\hspace*{0.3cm} \texttt{\textsf{else if (strong\_bond\_f (rev j) i =}} \texttt{\textsf{snd\_last\_bond\_of\_jun (rev j) i) $\land \ \sim$(i = 0)}}  \\
\hspace*{0.5cm} \texttt{\textsf{then snd\_last\_bond\_dir\_cond j i $\ast$}}  \\
\hspace*{1.0cm} \texttt{\textsf{snd\_last\_bond\_modulus\_cond (rev j) i $\ast$}} \texttt{\textsf{(fw\_jun\_e\_sum (rev j) i t + fwrd\_path j i t)}}  \\
\hspace*{0.5cm} \texttt{\textsf{else snd\_last\_bond\_modulus\_cond (rev j) i $\ast$}} \texttt{\textsf{bond\_flow (rev j) i (strong\_bond\_f (rev j) i) t}}}  \\
\hline
\end{tabular}
\caption{Formalized Parts of Definition~\ref{DEF:fw_path}}
\label{tab:fw_path}
\end{table}

\noindent It is possible for a system to have a BG representation with only one junction as shown in Figure~\ref{fig:bg_exp}. In this case, there is no causal path (backward or forward), except the last and second last bonds with the effort or flow variables. We model this scenario in \holl~as:
\begin{definition}
\label{DEF:unit_jun}
{
\textup{\texttt{\textsf{
$\vdash_{}$ $\forall$j i t. backwrd\_path\_e j i t =  \\
\hspace*{1.8cm} (if LENGTH j - 1 = 0 then snd\_last\_bond\_e j i t else backwrd\_path j i t) \\
$\vdash_{}$ $\forall$j i t. backwrd\_path\_f j i t =  \\
\hspace*{1.8cm}  (if LENGTH j - 1 = 0 then snd\_last\_bond\_f j i t else backwrd\_path j i t) \\
$\vdash_{}$  $\forall$j i t. final\_fwrd\_path\_e j i t =  \\
\hspace*{1.8cm}  (if LENGTH j - 1 = 0 then last\_bond\_e j i t else final\_fwrd\_path j i t)  \\
$\vdash_{}$  $\forall$j i t. final\_fwrd\_path\_f j i t =   \\
\hspace*{1.8cm} (if LENGTH j - 1 = 0 then last\_bond\_f j i t else final\_fwrd\_path j i t)
}}}}
\end{definition}

The functions \texttt{\textsf{backwrd\_path\_e}} and \texttt{\textsf{backwrd\_path\_f}} check the length of the junction list \texttt{\textsf{j}} and if it is equal to zero then there is no backward path except the second last bond with the effort and flow variables, respectively. If this condition is not true, then there is a causal path. Similarly, the functions \texttt{\textsf{final\_fwrd\_path\_e}} and \texttt{\textsf{final\_fwrd\_path\_f}} extract the effort and flow of the last bond in the case of only one junction, otherwise, there is a forward path to follow in a BG representation.
\par
Next, we model the causal paths as the following \holl~function:

\begin{definition}
\label{DEF:causal_paths}
{
\textup{\texttt{\textsf{
$\vdash_{}$  $\forall$j i t. causal\_paths j i t = (last\_bond\_dir j i $\ast$  \\
\hspace*{1.3cm} final\_fwrd\_path j i t + snd\_last\_bond\_dir j i $\ast$ backwrd\_path j i t)
}}}}
\end{definition}

A causal path consists of forward or backward path or both depending upon the causality, position and structure of the junction. For example,  in Figure~\ref{fig:novelbg}a, to obtain the summation equation (for flow variable) of Bond $9$, we follow both forward and backward paths by eliminating the last two bonds of $0\_junction$. It is important to note that the power direction of the skipped bonds is important in the law of summation. Therefore, we multiply the power direction of the last bond (\texttt{\textsf{last\_bond\_dir}}) with the forward path (\texttt{\textsf{final\_fwrd\_path}}) and add it to the backward path (\texttt{\textsf{backwrd\_path}}) multiplied with the power direction of the second last bond (\texttt{\textsf{snd\_last\_bond\_dir}}).

\begin{definition}
\label{DEF:Paths_selection_based_on_jun_num}
{
\textup{\texttt{\textsf{
$\vdash_{}$  $\forall$j i t. search\_path\_e j i t = (if i = 0 then (snd\_last\_bond\_dir j i $\ast$ \\
\hspace*{0.7cm}   snd\_last\_bond\_e j i t + last\_bond\_dir j i $\ast$ final\_fwrd\_path\_e j i t) \\
\hspace*{0.7cm}  else if (i = LENGTH j - 1) then (last\_bond\_dir j i $\ast$ last\_bond\_e j i t \\
\hspace*{2.5cm}  + snd\_last\_bond\_dir j i $\ast$ backwrd\_path\_e j i t) else Cx(\&0)) \\
$\vdash_{}$   $\forall$j i t. search\_path\_f j i t = (if i = 0 then (snd\_last\_bond\_dir j i $\ast$  \\
\hspace*{0.7cm}  snd\_last\_bond\_f j i t + last\_bond\_dir j i $\ast$ final\_fwrd\_path\_f j i t) \\
\hspace*{0.7cm}  else if (i = LENGTH j - 1) then (last\_bond\_dir j i $\ast$ last\_bond\_f j i t \\
\hspace*{2.5cm}  + snd\_last\_bond\_dir j i $\ast$ backwrd\_path\_f j i t) else Cx(\&0))
}}}}
\end{definition}

The functions \texttt{\textsf{search\_path\_e}} and \texttt{\textsf{search\_path\_f}} search a path for the effort and flow variables, respectively, based on the position (first or last) of a junction in a junction list \texttt{j}.

\begin{definition}
\label{DEF:Paths_selection_based_on_integral_causality}
{
\textup{\texttt{\textsf{
$\vdash_{}$  $\forall$j i t. path\_select\_t j i t =  \\
\hspace*{0.6cm}  (if (strong\_bond\_t j i = last\_bond\_of\_jun j i) $\land \ \sim$(i = LENGTH j - 1)  \\
\hspace*{1.7cm}  then final\_fwrd\_path j i t  \\
\hspace*{0.6cm}  else if (strong\_bond\_t j i = snd\_last\_bond\_of\_jun j i) $\land \ \sim$(i = 0)  \\
\hspace*{1.6cm}  then (backwrd\_path j i t) else bond\_effort j i (strong\_bond\_t j i) t) \\
$\vdash_{}$  $\forall$j i t. path\_select\_f j i t = (if (strong\_bond\_f j i = last\_bond\_of\_jun j i) $\land$ \\
\hspace*{0.6cm}   $\sim$(i = LENGTH j - 1) then final\_fwrd\_path j i t \\
\hspace*{0.6cm}  else if (strong\_bond\_f j i = snd\_last\_bond\_of\_jun j i) $\land \ \sim$(i = 0) \\
\hspace*{1.6cm}  then (backwrd\_path j i t) else bond\_flow j i (strong\_bond\_f j i) t)
}}}}
\end{definition}

The energy storing elements I and C exhibit integral causality, i.e., component C has a causality towards a junction, whereas I has a causality in the opposite direction, as provided in Table \ref{tbl:basic_components_table}.
The function \texttt{\textsf{path\_select\_t}} selects a path by matching a strong bond (\texttt{\textsf{strong\_bond\_t}}) with the last or second last bond of the junction \texttt{\textsf{i}} and its position in the list of junctions \texttt{\textsf{j}}. Similarly, the \texttt{\textsf{path\_select\_f}} chooses a path depending upon the strong bond (\texttt{\textsf{strong\_bond\_f}}), second last bond (\texttt{\textsf{snd\_last\_bond\_of\_jun}}), last bond (\texttt{\textsf{last\_bond\_of\_jun}}) of the junction \texttt{\textsf{i}} and the position of the junction.
\par
We have formalized the law of equality for a junction (0,1) in \holl, as presented in Definition~\ref{DEF:equality_laws}. Now, we formalize the equality law for the case, when a junction is following a causal path as follows:

\begin{definition}
\label{DEF:Paths_selection_based_on_jun_type}
{
\textup{\texttt{\textsf{
$\vdash_{}$  $\forall$j i k t. path\_selection j i k t =   \\
\hspace*{0.6cm}  (if type\_of\_jun j i = F then (bond\_effort j i k t = path\_select\_t j i t)  \\
\hspace*{0.7cm}  else (bond\_flow j i k t = path\_select\_f j i t))
}}}}
\end{definition}

The function \texttt{\textsf{path\_selection}} selects a path for equality laws based on the type of a junction, which can be $0$ or $1$.
\par
Similarly, we formalize equality law for the differential causal elements as follows:

\begin{definition}
\label{DEF:Paths_selection_for_diff_causality}
{
\textup{\texttt{\textsf{
$\vdash_{}$  $\forall$j i k t. path\_selection\_der j i k t = (if (type\_of\_jun j i = F) then \\
\hspace*{0.2cm}  ($\lambda$t. vector\_derivative (bond\_effort j i k) (at t)) =  \\
\hspace*{0.2cm}  $\lambda$t. vector\_derivative (path\_select\_t j i) (at t)) else  \\
\hspace*{0.2cm}  ($\lambda$t. vector\_derivative (bond\_flow j i k) (at t)) =  \\
\hspace*{0.2cm}  ($\lambda$t. vector\_derivative (path\_select\_f j i) (at t)))
}}}}
\end{definition}

The causal elements I and C exhibiting the differential causality are provided in Table~\ref{tbl:basic_components_table}.
The function \texttt{\textsf{path\_selection\_der}} models equality law by taking derivative on both sides of the equality law's equation, obtained using the type of junction and the causality of the strong bond.
\par
Now, we formalize the summation laws along with a causal path depending on the causality, type and position of a junction as follows:

\begin{definition}
\label{DEF:jun_sum_types}
{
\textup{\texttt{\textsf{
$\vdash_{}$  $\forall$j i t. middle\_jun\_sum j i t = (if type\_of\_jun j i = T  \\
\hspace*{0.5cm}  then jun\_1\_e j i t +  causal\_paths j i t else jun\_0\_f j i t +  causal\_paths j i t)  \\
$\vdash_{}$  $\forall$j i t. side\_jun\_sum j i t = (if (type\_of\_jun j i = T)   \\
\hspace*{0.5cm}  then jun\_1\_e j i t +  search\_path\_e j i t else jun\_0\_f j i t +  search\_path\_f j i t)
}}}}
\end{definition}

The function \texttt{\textsf{type\_of\_jun}} accepts a list of junctions \texttt{\textsf{j}}, an element \texttt{\textsf{i}} of the list \texttt{\textsf{j}} and returns \textbf{T} for a $1\_junction$, otherwise it returns \textbf{F}. The function  \texttt{\textsf{middle\_jun\_sum}} combines summation law for effort and flow variables to the Boolean values \textbf{T} and \textbf{F}, respectively, alongwith the causal paths (\texttt{\textsf{causal\_paths}}).
A junction ($0$ or $1$) attached with junctions on both sides or on one side only is known as middle or side junction, respectively.
The function \texttt{\textsf{side\_jun\_sum}} is quite similar to that of \texttt{\textsf{middle\_jun\_sum}} except the paths, \texttt{\textsf{search\_path\_e}} and \texttt{\textsf{search\_path\_f}}. These paths are selected based on the type of junction. In Figure~\ref{fig:novelbg}a, $0\_junction$ is representing a middle junction, while $1\_junctions$ on both sides of the BG representation are side junctions.

\begin{definition}
\label{DEF:jun_sum_based_on_position}
{
\textup{\texttt{\textsf{
$\vdash_{}$  $\forall$j i t. jun\_sum j i t =    \\
\hspace*{5.2cm}   then side\_jun\_sum j i t else middle\_jun\_sum j i t)
}}}}
\end{definition}

The summation law of a junction also depends on its position in a BG representation.  The function \texttt{\textsf{jun\_sum}} checks the position of a junction (first or last) and applies the summation law (\texttt{\textsf{side\_jun\_sum}}), otherwise it applies the summation law on the middle junction (\texttt{\textsf{middle\_jun\_sum}}).

\begin{definition}
\label{DEF:jun_sum_final}
{
\textup{\texttt{\textsf{
$\vdash_{}$  $\forall$j i t. jun\_sum\_final j i t = (jun\_sum j i t = Cx(\&0))
}}}}
\end{definition}

According to the summation law of a junction, the sum of all the effort or flow variables is equal to zero. Thus, the function \texttt{\textsf{jun\_sum\_final}} models the final form (summation equal to zero) of the summation law.

\begin{definition}
\label{DEF:jun_sum_based_on_diff_causality}
{
\textup{\texttt{\textsf{
$\vdash_{}$  $\forall$j i t. jun\_sum\_der j i t =   \\
\hspace*{0.2cm}  ($\lambda$t. vector\_derivative (jun\_sum j i) (at t)) = ($\lambda$t. vector\_derivative ($\lambda$t. vec 0) (at t))
}}}}
\end{definition}

The function \texttt{\textsf{jun\_sum\_der}} accepts a list of junctions \texttt{\textsf{j}}, an element \texttt{\textsf{i}} of the list \texttt{\textsf{j}} and a time variable \texttt{\textsf{t}}, and returns a differential form of the summation law.

\begin{definition}
\label{DEF:path_selection_based_on_free_causality}
{
\textup{\texttt{\textsf{
$\vdash_{}$  $\forall$j i k t. res\_path\_selection j i k t =    \\
\hspace*{0.3cm}  (if causality j i k = F then (if  type\_of\_jun j i = F  \\
\hspace*{0.3cm}  then bond\_effort j i k t = path\_select\_t j i t else jun\_sum\_final j i t)  \\
\hspace*{0.3cm}  else if type\_of\_jun j i = F  \\
\hspace*{0.3cm}  then jun\_sum\_final j i t else bond\_flow j i k t = path\_select\_f j i t)
}}}}
\end{definition}

The resistive component of a BG representation has a free causality, i.e., its causality follows a structure of junctions rather than that of its components.
The function \texttt{\textsf{res\_path\_selection}} covers all possible cases for the resistive element having free causality and returns both summation \texttt{\textsf{jun\_sum\_final}} and the equality law based on the causality of a resistive element and the type of a junction.

\begin{definition}
\label{DEF:branch_present}
{
\textup{\texttt{\textsf{
$\vdash_{}$   $\forall$j j1 p q p1 q1 t. branch\_presence j j1 p q p1 q1 t =  \\
\hspace*{0.3cm}  (if (branch j p q = T) $\land$ (branch j1 p1 q1 = T)  \\
\hspace*{0.5cm}  then (if branch\_num j p q = branch\_num j1 p1 q1  \\
\hspace*{1.4cm}  then if type\_of\_ele j1 p1 q1 = 6 then (bond\_effort j p q t =  \\
\hspace*{2.4cm}   EL 1 (modulus j1 p1 q1) $\ast$ bond\_flow j1 p1 q1 t) $\lor$  \\
\hspace*{0.5cm}   (bond\_flow j p q t = EL 0 (modulus j1 p1 q1) $\ast$ bond\_effort j1 p1 q1 t) $\lor$ \\
\hspace*{0.3cm}  else (bond\_effort j p q t = EL 0 (modulus j1 p1 q1) $\ast$ bond\_effort j1 p1 q1 t) $\lor$ \\
\hspace*{0.5cm}  (bond\_flow j p q t) = EL 1 (modulus j1 p1 q1) $\ast$ bond\_flow j1 p1 q1 t) else F) else F) \\
$\vdash_{}$  $\forall$ j j1 t. branch\_main j j1 t = (if j = [ ] $\lor$ j1 = [ ] then F \\
\hspace*{7.2cm}  else (branch\_jun j j1 (LENGTH j - 1) t))
}}}}
\end{definition}

The function \texttt{\textsf{branch\_presence}} accepts two different lists of junctions \texttt{\textsf{j,j1}}, a junction \texttt{\textsf{p}} of the list \texttt{\textsf{j}}, a bond \texttt{\textsf{p1}} of the junction \texttt{\textsf{j}}, a different junction \texttt{\textsf{q}} of the list \texttt{\textsf{j1}}, a bond \texttt{\textsf{q1}} of the junction \texttt{\textsf{j1}} and a time variable \texttt{\textsf{t}}, and returns a set of equations capturing the equality laws (for effort and flow) for common bonds of \texttt{\textsf{j}} and \texttt{\textsf{j1}}. Here \texttt{\textsf{j}} and \texttt{\textsf{j1}} represent parent and child branches, respectively. The presence of a branch \texttt{\textsf{branch}} in a BG representation is associated with Boolean value \textbf{T} in the $6$-tuple. In a BG representation, multiple branches may appear on a junction, so in order to distinguish between them, we have assigned a number \texttt{\textsf{branch\_num}} to them as described in Section~\ref{SEC:flow of bg}. The connecting/common bond of two branches can be a two-port element (TF, GY) or a simple power bond. The function \texttt{\textsf{type\_of\_ele}} returns a Boolean value \textbf{T}, if the connecting bond is a gyrator (GY) and provides the respective equations (effort and flow) with the modulus of GY, otherwise, it returns \textbf{F} and provides respective equations (effort and flow) with the modulus of transformer (TF) or a connecting bond (with modulus $1$). For example, in Figure~\ref{fig:novelbg}b, Bond $131$ or $212$ is a common bond connecting $0\_junction$ to $1\_junction$.
The function \texttt{\textsf{branch\_main}} accepts two list of junctions \texttt{\textsf{j,j1}}, which are connected together, and a time variable \texttt{\textsf{t}}, and returns a boolean \textbf{F}, if the lists \texttt{\textsf{j}} or \texttt{\textsf{j1}} are empty. Otherwise, the function \texttt{\textsf{branch\_jun}} checks all the junctions of list \texttt{\textsf{j}} for branch presence.

\begin{definition}
\label{DEF:loop_present}
{
\textup{\texttt{\textsf{
$\vdash_{}$  $\forall$j i k i1 n t. loop\_presence j i k i1 n t =   \\
\hspace*{0.3cm}   (if branch j i1 n = T then (if branch\_num j i k = branch\_num j i1 n  \\
\hspace*{1.2cm}   then if type\_of\_ele j i1 n = 6 then (bond\_effort j i k t =  \\
\hspace*{2cm}   EL 1 (modulus j i1 n) $\ast$ bond\_flow j i1 n t) $\lor$  \\
\hspace*{2cm}   (bond\_flow j i k t = EL 0 (modulus j i1 n) $\ast$ bond\_effort j i1 n t)  \\
\hspace*{0.4cm}   else (bond\_effort j i k t = EL 0 (modulus j i1 n $\ast$ bond\_effort j i1 n t) $\lor$   \\
\hspace*{0.4cm}   bond\_flow j i k t = EL 1 (modulus j i1 n) $\ast$ bond\_flow j i1 n t) else F) else F)
}}}}
\end{definition}

The function \texttt{\textsf{loop\_presence}} accepts a list of junctions \texttt{\textsf{j}}, a junction \texttt{\textsf{i}} of the list \texttt{\textsf{j}}, a bond \texttt{\textsf{k}} of the junction \texttt{\textsf{i}}, a junction \texttt{\textsf{i1}} of the list \texttt{\textsf{j}}, a bond \texttt{\textsf{n}} of the junction \texttt{\textsf{i1}} and a time variable \texttt{\textsf{t}}, and provides a set of equations capturing the equality laws (for effort and flow) for common bonds. Firstly, the function  \texttt{\textsf{loop\_presence}} checks the presence of a branch on the junction \texttt{\textsf{i1}}, matches the branch numbers of both junctions \texttt{\textsf{i}} and \texttt{\textsf{i1}} (to check that they are connected together or not) and finally checks the presence of the gyrator component (integer value $6$). If the branch numbers of both junctions do not match or there is no branch then the returned value is \textbf{F}.

\begin{definition}
\label{DEF:recur_fun_for_loop_bonds}
{
\textup{\texttt{\textsf{
$\vdash_{}$  $\forall$j i k i1 n t. loop\_bonds j i k i1 0 t =  \\
\hspace*{0.3cm}  (if 0 = k then F else loop\_presence j i k i1 0 t) $\land$  \\
\hspace*{0.4cm}   loop\_bonds j i k i1 (SUC n) t = (if SUC n = k then F  \\
\hspace*{0.5cm}   else (loop\_presence j i k i1 (SUC n) t)$\lor$ (loop\_bonds j i k i1 n t))  \\
$\vdash_{}$   $\forall$j i k m t. loop\_bonds\_lst j i k i1 t = loop\_bonds j i k i1 (bonds\_length j i1 - 1) t
}}}}
\end{definition}

The function \texttt{\textsf{loop\_bonds}} ensures that if a loop (a branch with both ends connected to junctions) exists on any junction then both of its starting and the ending bonds (\texttt{\textsf{k, n}}) do not have similar bond numbers, i.e, to distinguish both connecting bonds of the loop. The function \texttt{\textsf{loop\_bonds\_lst}} accepts a list of junctions \texttt{\textsf{j}}, a junction \texttt{\textsf{i}} of the list \texttt{\textsf{j}}, a bond \texttt{\textsf{k}} of the junction \texttt{\textsf{i}}, a junction \texttt{\textsf{i1}} from the list \texttt{\textsf{j}} and ensures the presence of a loop in a junction \texttt{\textsf{i1}} by checking all of its bonds using (\texttt{\textsf{bonds\_length}}).

\begin{definition}
\label{DEF:recur_fun_for_loop_jun}
{
\textup{\texttt{\textsf{
$\vdash_{}$  $\forall$j i m i1 t. loop\_jun j i k 0 t = loop\_bonds\_lst j i k 0 t $\land$  \\
\hspace*{0.3cm}  loop\_jun j i k (SUC i1) t = (loop\_bonds\_lst j i k (SUC i1) t) $\lor$ (loop\_jun j i k i1 t)   \\
$\vdash_{}$   $\forall$j i k t.loop\_jun\_lst j i m t = loop\_jun j i k (LENGTH j - 1) t
}}}}
\end{definition}

A causal loop is a closed causal path with bonds either connected to a similar junction or two different junctions as described in Table~\ref{tab:defs}.
The function \texttt{\textsf{loop\_jun}} ensures the presence of a loop in a junction \texttt{\textsf{i1}} by checking junctions of the \texttt{\textsf{j}} recursively. The function \texttt{\textsf{loop\_jun\_lst}} checks all the junctions of the list \texttt{\textsf{j}} to find the junction on which the connecting bond of the loop is attached.

\begin{definition}
\label{DEF:causal_loop}
{
\textup{\texttt{\textsf{
$\vdash_{}$  $\forall$j i k t. causal\_loop j i k t = (if (branch j i k = T) $\land$   \\
\hspace*{4.5cm}  $\sim$(branch\_num j i k = 0) then (loop\_jun\_lst j i k t) else F
}}}}
\end{definition}

A \texttt{\textsf{causal\_loop}} ensures the presence of a loop in a BG representation and provides its equations by using the function \texttt{\textsf{loop\_jun\_lst}}.

\subsection{Formalization of the Case Selection} \label{sec:caseselect}

Here, we define some cases of a BG representation based on our formalization till now.

\begin{definition}
\label{DEF:cases}
{
\textup{\texttt{\textsf{
$\vdash_{}$  $\forall$j i t. frst\_ordr\_ele\_a j i t = jun\_sum\_final j i t \\
$\vdash_{}$  $\forall$j i k t. frst\_ordr\_ele\_b j i k t = path\_selection j i k t \\
$\vdash_{}$  $\forall$j i k t. zero\_ordr\_ele j i k t = res\_path\_selection j i k t \\
$\vdash_{}$  $\forall$j i t. diff\_causal\_ele\_a j i t = jun\_sum\_der j i t  \\
$\vdash_{}$  $\forall$j i k t. diff\_causal\_ele\_b j i k t = path\_selection\_der j i k t  \\
$\vdash_{}$  $\forall$j i k t. branch\_type\_a j i k t = (jun\_sum\_final j i t) $\lor$ (causal\_loop j i k t)  \\
$\vdash_{}$  $\forall$j i k t. branch\_type\_b j i k t = (path\_selection j i k t) $\lor$ (causal\_loop j i k t)
}}}}
\end{definition}

The functions \texttt{\textsf{frst\_ordr\_ele\_a}} and \texttt{\textsf{frst\_ordr\_ele\_b}} model all those cases, where the junction, components and causality appear in such a way that the summation and equality laws are deducted, respectively. For example, in Figure~\ref{fig:novelbg}b, the bond of the $1\_junctions$ with a component I (having integral causality) results into summation law and the bond of $1\_junction$ connected with component C (having integral causality) results into equality law by following causal paths. The function \texttt{\textsf{zero\_ordr\_ele}} models junctions with the resistive component (R). The function \texttt{\textsf{diff\_causal\_ele\_a}} is similar to the first order case except the elements with differential causality. These functions provide differential equations of the BG representation. Lastly, the function \texttt{\textsf{branch\_type\_a}} is also similar to first order case except the presence of a causal loop. The detailed definition can be found in our \holl~proof script~\cite{formalizationbg} for detailed cases of a BG representation.

\begin{definition}
\label{DEF:case_select}
{
\textup{\texttt{\textsf{
$\vdash_{}$  $\forall$j i k t. case\_selection j i k t =   \\
\hspace*{0.3cm} (if P1 then frst\_ordr\_ele\_a j i t else if P2 then frst\_ordr\_ele\_b j i k t   \\
\hspace*{1.3cm}  else if P3 then zero\_ordr\_ele j i k t else if P4 then diff\_causal\_ele\_a j i t   \\
\hspace*{1.9cm}  else if P5 then branch\_type\_a j i k t else if P6 then T else F)
}}}}
\end{definition}

\newcolumntype{P}[1]{>{\centering\arraybackslash}p{#1}}
\newcolumntype{M}[1]{>{\centering\arraybackslash}m{#1}}
\begin{table}[!ht]
\centering
\footnotesize
\renewcommand{\arraystretch}{1.01}
\begin{tabular}{|M{0.6cm} |M{14.8cm}|}
\hline
\textbf{Part} & \textbf{Formalized Form}\\
\hline
\texttt{\textsf{P1}} & {\raggedright  \texttt{\textsf{(type\_of\_jun j i = T) $\land$ (type\_of\_ele j i k = 2) $\land$}}  \\
\hspace*{0.2cm} \texttt{\textsf{(causality j i k = F) $\lor$ (type\_of\_jun j i = F) $\land$}}
 \texttt{\textsf{(type\_of\_ele j i k = 3) $\land$ (causality j i k = T)}}} \\
\hline
\texttt{\textsf{P2}} & {\raggedright  \texttt{\textsf{(type\_of\_jun j i = T) $\land$ (type\_of\_ele j i k = 3) $\land$}}  \\
\hspace*{0.2cm} \texttt{\textsf{(causality j i k = T) $\lor$ (type\_of\_jun j i = F) $\land$}}
  \texttt{\textsf{(type\_of\_ele j i k = 2) $\land$ (causality j i k = F)}}}  \\
\hline
\texttt{\textsf{P3}} & {\raggedright  \texttt{\textsf{(type\_of\_jun j i = T) $\land$ (type\_of\_ele j i k = 4) $\land$}}
 \texttt{\textsf{(causality j i k = T) $\lor$ (type\_of\_jun j i = F) $\land$}} \\
\hspace*{-0.1cm} \texttt{\textsf{(type\_of\_ele j i k = 4) $\land$ (causality j i k = F) $\lor$}}
 \texttt{\textsf{(type\_of\_jun j i = T) $\land$ (type\_of\_ele j i k = 4) $\land$}}  \\
\hspace*{0.2cm} \texttt{\textsf{(causality j i k = F) $\lor$ (type\_of\_jun j i = F) $\land$}}
  \texttt{\textsf{(type\_of\_ele j i k = 4) $\land$ (causality j i k = T)}}}  \\
\hline
\texttt{\textsf{P4}} & {\raggedright  \texttt{\textsf{(type\_of\_jun j i = T) $\land$ (type\_of\_ele j i k = 3) $\land$}}  \\
\hspace*{0.2cm} \texttt{\textsf{(causality j i k = F) $\lor$ (type\_of\_jun j i = F) $\land$}}
 \texttt{\textsf{(type\_of\_ele j i k = 2) $\land$ (causality j i k = T)}}}  \\
\hline
\texttt{\textsf{P5}} & {\raggedright   \texttt{\textsf{((type\_of\_jun j i = T) $\land$ (branch j i k = T) $\land$ }}
  \texttt{\textsf{(causality j i k = F)) $\land$ (type\_of\_ele j i k = 0 $\lor$}}  \\
\hspace*{0.2cm} \texttt{\textsf{type\_of\_ele j i k = 5 $\lor$ type\_of\_ele j i k = 6) $\lor$}}
  \texttt{\textsf{((type\_of\_jun j i = F) $\land$ (branch j i k) = T) $\land$}}  \\
\hspace*{-0.1cm} \texttt{\textsf{(causality j i k) = T)) $\land$ (type\_of\_ele j i k = 0 $\lor$}}
 \texttt{\textsf{type\_of\_ele j i k = 5 $\lor$ type\_of\_ele j i k = 6)}}}  \\
\hline
\texttt{\textsf{P6}} & {\raggedright  \texttt{\textsf{(type\_of\_ele j i k = 0 $\lor$ type\_of\_ele j i k = 1 $\lor$}}
  \texttt{\textsf{type\_of\_ele j i k = 5 $\lor$ type\_of\_ele j i k = 6)}}} \\
\hline
\end{tabular}
\caption{Formalized Parts of Definition~\ref{DEF:case_select}}
\label{tab:case_select}
\end{table}

The \texttt{\textsf{case\_selection}} takes a list of junctions \texttt{\textsf{j}}, an element \texttt{\textsf{i}} of the list \texttt{\textsf{j}}, a bond number \texttt{\textsf{k}} and a time variable \texttt{\textsf{t}} and provides a suitable case for a junction depending on the causality of bonds, type of elements and the type of junction. The detailed formalized form of the parts \texttt{\textsf{P1, P2, P3}} and \texttt{\textsf{P4}} is given in the Table~\ref{tab:case_select}.

\begin{definition}
\label{DEF:bond_select}
{
\textup{\texttt{\textsf{
$\vdash_{}$   $\forall$j i k t. (bond\_selection j i 0 t = case\_selection j i 0 t) $\land$   \\
\hspace*{0.2cm}   bond\_selection j i (SUC k) t = (case\_selection j i (SUC k) t) $\land$ (bond\_selection j i k t)   \\
$\vdash_{}$   $\forall$j i k t. bond\_selection\_lst j i t = bond\_selection j i (bonds\_length j i - 1) t
}}}}
\end{definition}

The function \texttt{\textsf{bond\_selection}} is a recursive function, which applies the function \texttt{\textsf{case\_selection}} to the bonds of a junction. Similarly, the function \texttt{\textsf{bond\_selection\_lst}} applies case selection to all the bonds of a junction.

\begin{definition}
\label{DEF:jun_select}
{
\textup{\texttt{\textsf{
$\vdash_{}$  $\forall$j i t. jun\_selection j i t = bond\_selection\_lst j 0 t $\land$  \\
\hspace*{0.3cm}   jun\_selection j (SUC i) t = (bond\_selection\_lst j (SUC i) t) $\land$ (jun\_selection j i t)  \\
$\vdash_{}$   $\forall$j t. bg\_main j t = (if j = [ ]  then F else (jun\_selection j (LENGTH j - 1) t))
}}}}
\end{definition}

The function \texttt{\textsf{jun\_selection}} is a recursive function, which applies the function \texttt{\textsf{bond\_selection\_lst}} to all the bonds of the junction \texttt{\textsf{i}} of the list \texttt{\textsf{j}}.
The function \texttt{\textsf{bg\_main}} is the main function of BG formalization that accepts only two parameters, i.e., a list of junctions \texttt{\textsf{j}} and time variable \texttt{\textsf{t}}, and returns all the required equations of a BG representation by applying the function \texttt{\textsf{jun\_selection}} on all the junctions of the list \texttt{\textsf{j}}.


\subsection{Formalization of the State-space Model} \label{sec:ssrep}

Finally, the state-space model for a BG representation is formalized in \holl~as:

\begin{definition}
\label{DEF:ss_model}
{
\textup{\texttt{\textsf{
$\vdash_{}$  $\forall$A B x x\_der u. ss\_model A B x x\_der u =   \\
\hspace*{9.6cm}   (x\_der = A $\ast \ast$ x + B $\ast \ast$ u)
}}}}
\end{definition}

The function \texttt{\textsf{ss\_model}} accepts a system matrix \texttt{\textsf{A:$\mathbb{C}^{N\times N}$}}, input matrix \texttt{\textsf{B:$\mathbb{C}^{P\times N}$}}, state vector \texttt{\textsf{x:$\mathbb{C}^{N}$}}, derivative of state vector with respect to time \texttt{\textsf{x\_der:$\mathbb{C}^{N}$}} and input vector\texttt{\textsf{u:$\mathbb{C}^{P}$}}, and provides a state-space model for a BG representation.


\section{Formal Verification of the Bond Graphs Properties}\label{SEC:bond_graphs_properties}

Stability is an important control characteristic of physical systems that dampens out any oscillation in the performance of the system caused by various disturbances, and thus restores systems to the equilibrium conditions~\cite{nise2007control}. Thus, a stable system provides a stable response/output to a bounded input. The BG representation of a system is expressed as state-space models, which are based on vectors and matrices, in particular the system matrix. Stability of a system depends on the location of the poles in a complex plane, which are the eigenvalues of the system's matrix. Based on the location of the poles in a complex plane, any system can be categorized as of three types, namely, stable, marginally stable and unstable system. For a stable system, the eigenvalues of the system's matrix lie in the left half of the complex plane.

\begin{figure}[!ht]
\centering
\resizebox{0.5\hsize}{0.32\textheight}{\includegraphics{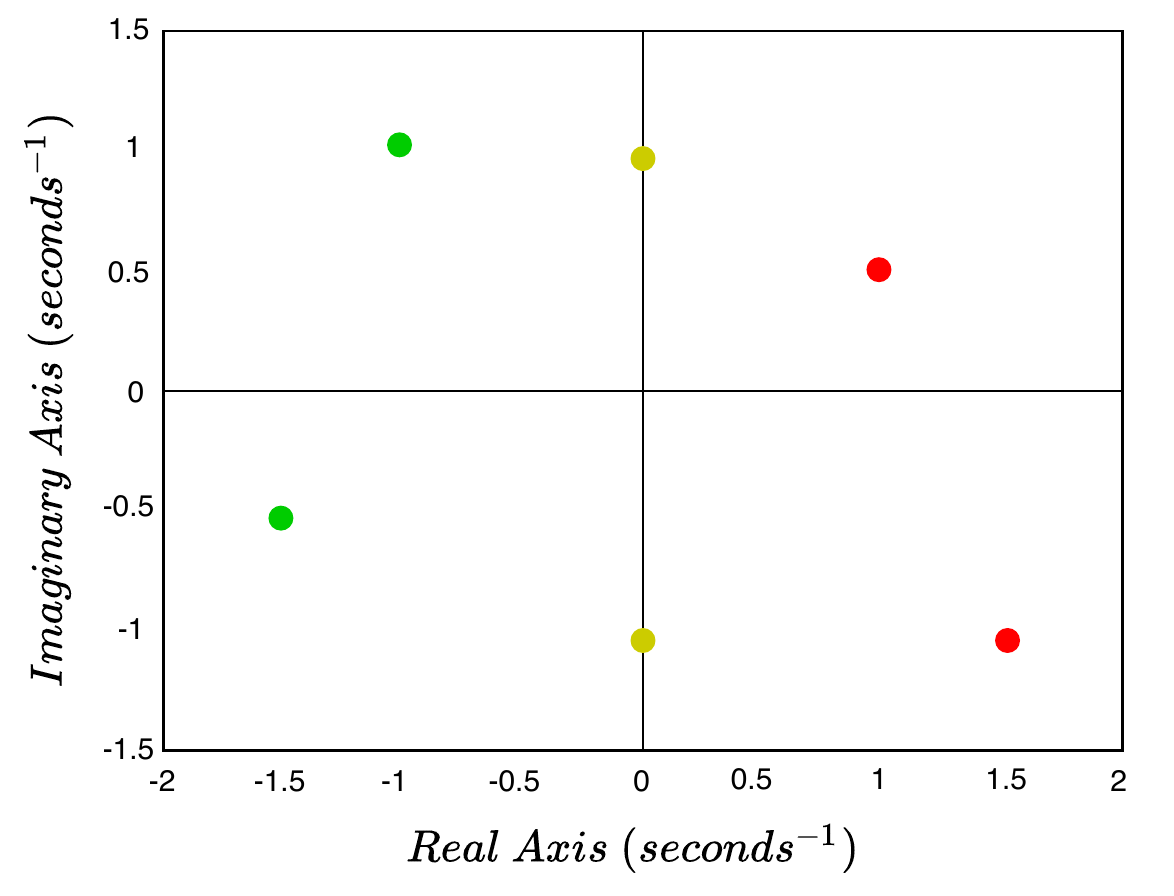}}
\caption{Stability Regions}
\label{fig:stable_region}
\end{figure}

\noindent In Figure~\ref{fig:stable_region}, green dots represent eigenvalues of a stable system.
Similarly, for an unstable system, the eigenvalues of the system's matrix lie in the right half of the complex plane, which are represented as red dots in Figure~\ref{fig:stable_region}. Moreover, for a marginally stable system, the eigenvalues of the system's matrix lie on the imaginary axis of the complex plane. Yellow dots (eigenvalues) exactly lie on the imaginary axis representing a marginally stable system as shown in Figure~\ref{fig:stable_region}.
\par
\noindent For a system matrix A, the characteristic equation is described as:
\[Ax = cx\]
\noindent Where x is the eigenvector and c represents the eigenvalues.
We can find the eigenvalues of the system matrix A by solving the following equations involving the determinant of a matrix.
\[|A -cI| = 0\]
\noindent Where I is an identity matrix.
Stability has been formalized in \holl~for the case of polynomial~\cite{ahmed2018formal}. However, it cannot incorporate the stability of the state-space models. We model the notion of stability, incorporating the state-space models, in \holl~as follows:

\begin{definition}
\label{DEF:stability_in_hol}
{
\textup{\texttt{\textsf{
$\vdash_{}$  $\forall$A. stable\_sys A =  \\
\hspace*{0.5cm} $\sim$ (\{c | cdet (A - c \%\%\% cmat (Cx (\&1))) = Cx (\&0) $\land$ Re (c) < \&0\} = EMPTY)  \\
$\vdash_{}$    $\forall$A. unstable\_sys A =  \\
\hspace*{0.5cm}  $\sim$ (\{c | cdet (A - c \%\%\% cmat (Cx (\&1))) = Cx (\&0) $\land$ Re (c) > \&0\} = EMPTY)  \\
$\vdash_{}$   $\forall$A. marginally\_stable\_sys A =  \\
\hspace*{0.5cm}   $\sim$ (\{c | cdet (A - c \%\%\% cmat (Cx (\&1))) = Cx (\&0) $\land$ Re (c) = \&0\} = EMPTY)
}}}}
\end{definition}

The function \texttt{\textsf{stable\_sys:}}$\mathbb{C}^{N\times N} \to$ Bool accepts a complex-valued matrix \texttt{\textsf{A}} and returns a Boolean value (\texttt{\textsf{T}}), if both conditions are satisfied, i.e., the first condition provides the eigenvalues of the state-space matrix \texttt{\textsf{A}}, whereas, the second condition ensures that the real part of the eigenvalues (roots) lie in left half of the complex plane and thus ensures a stable system. Here, \texttt{\textsf{cdet}} models the determinant of a complex-valued matrix. Similarly, \texttt{\textsf{cmat}} provides a complex-valued identity matrix. Moreover, the operator \texttt{\textsf{\%\%\%:$\mathbb{C} \to \mathbb{C}^{N\times M} \to \mathbb{C}^{N\times M}$}}, provides a scalar multiplication of a complex-valued matrix. Similarly, the functions \texttt{\textsf{unstable\_sys}} and \texttt{\textsf{marginally\_stable\_sys}} model an unstable and a marginally stable system, respectively.
\par
In the state-space representation, we deal with matrices, vectors and their corresponding operations.  For the stability of a system, entries of the matrix \texttt{\textsf{A}} are analyzed. Since, there are different arithmetic operations (+, $\ast$) involved and the parameters of the system matrix contain round-off values in our case~\cite{saeed2019comprehensive} so, rather than taking particular values of the parameters, we took general values and verified the theorems for (stable, unstable, marginally stable) square matrices of dimension $2$ and $3$.

\begin{theorem}
\label{THM:stable_mat}
{
\textup{\texttt{\textsf{
$\vdash_{}$   $\forall$a11 a12 a13 a21 a22 a23 a31 a32 a33 b1 c1 d1 r.   \\
\hspace*{0.3cm}  Cx (-- a33 - a22 - a11)  = Cx b1 +  Cx r $\land$ Cx (-- a13 $\ast$ a31 - a12 $\ast$ a21 - a23 $\ast$ a32 +  \\
\hspace*{1.0cm} a22 $\ast$ a33 + a11 $\ast$ a33 + a11 $\ast$ a22) = Cx c1 + Cx (b1 $\ast$ r) $\land$  \\
\hspace*{0.3cm} Cx (-- a11 $\ast$ a22 $\ast$ a33 - a12 $\ast$ a31 $\ast$ a23 - a13 $\ast$ a21 $\ast$ a32 + a11 $\ast$ a23 $\ast$   \\
\hspace*{1.7cm} a32 + a12 $\ast$ a21 $\ast$ a33 + a13 $\ast$ a31 $\ast$ a22) = Cx (c1 $\ast$ r) $\land$  \\
 \big(\&0 < r $\lor$ \big(\&0 < b1 $\land$ \big(b1 pow 2 - \&4 $\ast$ c1 < \&0 $\lor$ b1 pow 2 - \&4 $\ast$ c1 = \&0\big) $ \lor $  \\
\hspace*{0.3cm} \big(\&0 < b1 pow 2 - \&4 $\ast$ c1 $\land$ \big((b1 pow 2 - \&4 $\ast$ c1) < b1 $\lor$   \\
\hspace*{0.3cm} -- b1 < $\sqrt {\texttt{(b1 pow 2 - \&4 $\ast$ c1)}}$ \big)\big)\big)\big)
 $\Rightarrow$ stable\_sys
  $\begin{bmatrix} \texttt{\textsf{Cx\ a11}}\ \ \ \texttt{\textsf{Cx\ a12}}\ \ \ \texttt{\textsf{Cx\ a13}}  \\
\ \texttt{\textsf{Cx\ a21}}\ \ \ \texttt{\textsf{Cx\ a22}}\ \ \ \texttt{\textsf{Cx\ a23}} \\
 \texttt{\textsf{Cx\ a31}}\ \ \ \texttt{\textsf{Cx\ a32}}\ \ \ \texttt{\textsf{Cx\ a33}}  \end{bmatrix}$
}}}}
\end{theorem}

The above theorem provides all possible conditions on the entries of a matrix with dimension \texttt{\textsf{3$\times$3}} to be stable. This theorem is formally verified using complex matrix theory, which is developed as part of this work~\cite{formalizationbg}~and multivariate complex and real analysis theories available in the library of the \holl~theorem prover.
We used lemmas in the formalization of matrix stability to solve characteristics polynomial of the given matrix. These lemmas can be found in our \holl~proof script~\cite{formalizationbg}.
\par
Next, we verify some important properties of the state-space models of the given BG representations, when they represent some specific matrices, such as, upper triangular, lower triangular and diagonal matrices~\cite{ricardo2009modern,mirsky2012introduction}. We formalize these matrices as follows:

\begin{definition}
\label{DEF:Matrices_Def}
{
\textup{\texttt{\textsf{
$\vdash_{}$  $\forall$A. ut\_cmatrix A = (!i j.1 $ \le $ i $ \land $ i $ \le $ dimindex(:M) $ \land $  \\
\hspace*{3.9cm}  1 $\le$ j $\land$ j $ \le$ dimindex(:N) $ \land $ (j < i) $\Rightarrow$ (A$ \$ i \$ j $ = Cx(\&0))  \\
$\vdash_{}$   $\forall$A. lt\_cmatrix A = (!i j.1 $ \le $ i $ \land $ i $ \le $ dimindex(:M) $ \land $   \\
\hspace*{3.9cm}  1 $ \le $ j $ \land $ j $ \le $ dimindex(:N) $ \land $ (i < j) $\Rightarrow $  (A$ \$ i \$ j $ = Cx (\&0))   \\
$\vdash_{}$   $\forall$A. diagonal\_cmatrix A = !i j.1 $ \le $ i $ \land $ i $ \le $ dimindex(:M) $ \land $ \\
\hspace*{3.9cm}  1 $ \le $ j $ \land $ j $ \le $ dimindex(:N) $ \land $ $\sim$ (i = j) $\Rightarrow$ (A$ \$ i \$ j$= Cx (\&0))
}}}}
\end{definition}

The function \texttt{\textsf{ut\_cmatrix}} models a complex-valued upper triangular matrix, which is a square matrix with all its entries below the main diagonal equal to zero. Similarly, a square matrix whose entries above the main diagonal are zero is called the lower triangular matrix and is formalized in \holl~as a function \texttt{\textsf{lt\_matrix}}.
The function \texttt{\textsf{diagonal\_cmatrix}} provides a diagonal matrix, which is a special case of triangular matrices and has all its entries, other than diagonal, equal to zero.

\newcolumntype{P}[1]{>{\centering\arraybackslash}p{#1}}
\newcolumntype{M}[1]{>{\centering\arraybackslash}m{#1}}
\begin{table}[!ht]
\footnotesize
\renewcommand{\arraystretch}{1.1}
\begin{tabular}{|M{3.2cm}|M{12.4cm}|}
\hline
\textbf{Name} & \textbf{Formalized Form} \\
\hline
\texttt{Stable Matrix Property} & {\raggedright   {
\textup{\texttt{\textsf{
$\vdash_{thm}$   $\forall$A. \textbf{[A1]} (diagonal\_cmatrix A $\lor$ lt\_cmatrix A $\lor$ ut\_cmatrix A) $\land $  \\
\hspace*{-1.2cm}\textbf{[A2]} (stable\_diagonal\_ele\_cond A) $\Rightarrow$ stable\_sys A
}}}}
} \\
\hline
\texttt{Unstable Matrix Property} & {\raggedright  {
\textup{\texttt{\textsf{
$\vdash_{thm}$  $\forall$A. \textbf{[A1]} (diagonal\_cmatrix A $\lor$ lt\_cmatrix A $\lor$ ut\_cmatrix A) $\land $  \\
\hspace*{-0.45cm}\textbf{[A2]} (unstable\_diagonal\_ele\_cond A) $\Rightarrow$ unstable\_sys A
}}}}
} \\
\hline
\texttt{Marginally Stable Matrix Property} & {\raggedright  {
\textup{\texttt{\textsf{
$\vdash_{thm}$  $\forall$A. \textbf{[A1]} (diagonal\_cmatrix A $\lor$ lt\_cmatrix A $\lor$ ut\_cmatrix A) $\land $ \\
\hspace*{1.55cm}\textbf{[A2]} (marg\_stable\_diagonal\_ele\_cond A) $\Rightarrow$ marginally\_stable\_sys A
}}}}
} \\
\hline
\end{tabular}
\caption{Properties of Complex Matrices}
\label{tab:stab_mat_prop}
\end{table}

Table~\ref{tab:stab_mat_prop} presents formally verified properties of the complex matrices, providing stability, unstability and marginal stability under some conditions and the verification of these theorems are mainly based on the properties of complex-valued matrices along with some complex arithmetic reasoning.


\section{Application: Anthropomorphic Mechatronic Prosthetic Hand} \label{SEC:app}

\begin{figure}[!ht]
\begin{center}
 \begin{tabular}{cc}
  \resizebox{0.1\textwidth}{!}{\includegraphics{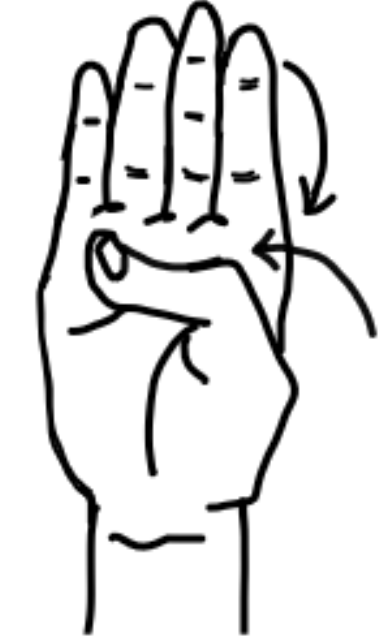}}
  \resizebox{0.1\textwidth}{!}{\includegraphics{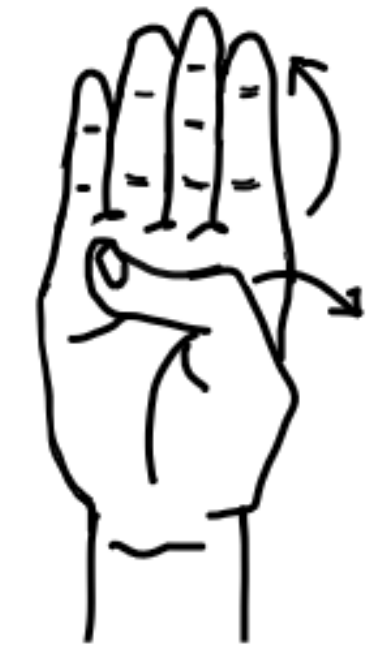}}
   \resizebox{0.09\textwidth}{!}{\includegraphics{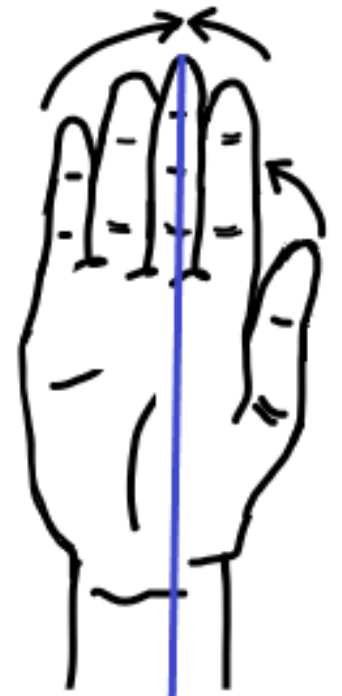}}
   \resizebox{0.1\textwidth}{!}{\includegraphics{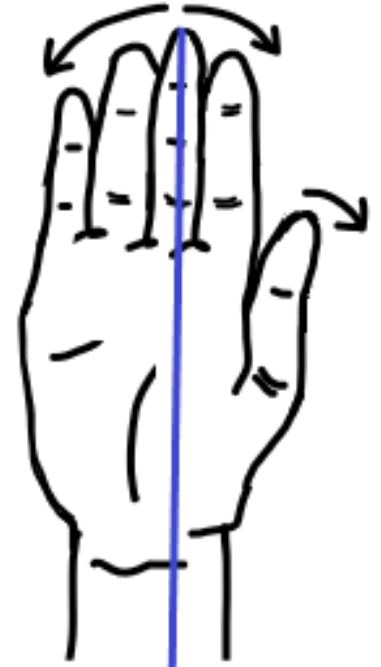}}\\
     {\small (a) Flexion}
      \hspace{3mm} {\small (b) Extension}
        \hspace{0.3mm} {\small (c) Adduction}
      \hspace{0.3mm}{\small (d) Abduction}
\end{tabular}
\end{center}
\caption{Movements of a Human Hand's Fingers}
\label{fig:human_hand}
\end{figure}

Anthropomorphic Mechatronic Hand~\cite{saeed2019comprehensive} is a robotic hand consisting of electrical and mechanical components. It  is capable of conducting movements of a human hand and is widely used in robotics, such as industrial, service, surgical robots~\cite{gaiser2008new,rhee2004door,honarpardaz2017finger}, etc. A prosthetic robotic hand improves the lives of the handicapped individuals by restoring the functions of the missing body parts. The accuracy and stability of such systems are crucial to replicate the desired movements of a human hand. A human hand consists of four fingers and a thumb involving flexion (flex.), extension (ext.), abduction (abd.), adduction (add.), up and down movements as shown in Figure~\ref{fig:human_hand}.  The flexion and extension are the movements of thumb and fingers, i.e., moving the tip of the finger/thumb towards and away from the palm as shown in Figure~\ref{fig:human_hand}, respectively.

Similarly, adduction and abduction move the fingers/thumb towards and away from the middle finger, respectively. Lastly, the up and down are the movements of the thumb~\cite{wang2017two}.

\begin{figure}[!ht]
	\centering
	\resizebox{0.61\hsize}{!}{\includegraphics{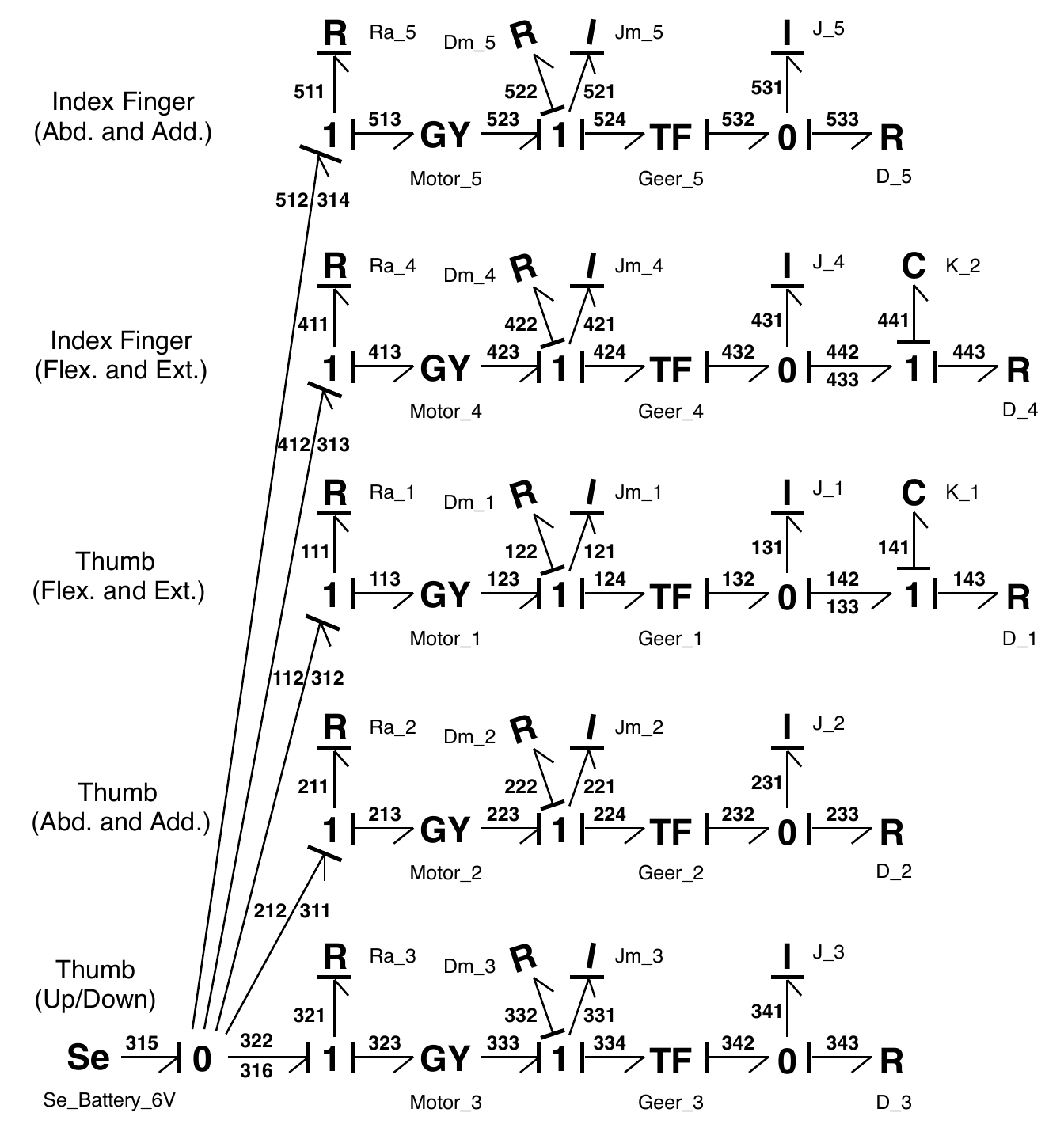}}
	\caption{Bond Graph of Anthropomorphic Mechatronic Prosthetic Hand}
	\label{fig:robotic_hand}
\end{figure}

\noindent A robotic hand replicates the structure of human hand consisting of bones, joints and muscles  by using frames, pulley-string mechanisms and electrical actuators, respectively. We formally analyzed all the movements of fingers of an anthropomorphic prosthetic hand~\cite{saeed2019comprehensive} by applying our formalization of BG, as described in Section~\ref{SEC:Formalization of BG model}. In this paper, we only present formal analysis of flexion and extension movements of the index finger.\\
A BG representation of an anthropomorphic mechatronic prosthetic hand is presented in Figure~\ref{fig:robotic_hand}. It consists of BG models of index finger for the case of abduction and adduction, and flexion and extension movements. It also contains BG models of thumb for the flexion and extension, abduction and adduction, and up and down movements. We have formalized each of these index finger and thumb movements using our formalization of the BG representation, presented in Section~\ref{SEC:Formalization of BG model}. \\
In this paper, we only present the formalization of flexion and extension movements of the index finger. In this regard, we model the index finger movements namely flexion (flex.) and extension (ext.) in \holl~as follows:

\begin{definition}
\label{DEF:index_fin_flex_ext}
{
\textup{\texttt{\textsf{
$\vdash_{}$  $\forall$e411 e412 f412 e413 f413 e421 f422 e423 f423 e424 f424    \\
  f433 f441 e442 f442 f443 Ra\_4 Motor\_4 Jm\_4 Dm\_4 Geer\_4 J\_4 K\_2 D\_4 p\_0 q\_0.   \\
\hspace*{-0.1cm}    index\_finger\_flex\_ext e411 e412 f412 e413 f413 e421 f422 e423 f423 e424 f424 e431 e432 f432    \\
\hspace*{-0.1cm}   e433 f433 f441 e442 f442 f443 Ra\_4 Motor\_4 Jm\_4 Dm\_4 Geer\_4 J\_4 K\_2 D\_4 p\_0 q\_0    \\
\hspace*{0.0cm}   = [0,T,[F,F,(F,0),4,[Cx(\&1); Cx(\&1)],(e411,res\_f Ra\_4 e411); T,T,(T,313),0,  \\
\hspace*{0.2cm}   [Cx (\&1); Cx (\&1)],(e412,f412); T,F,(F,0),6,[Cx \Big($\dfrac{\texttt{\&1}}{\texttt{Motor\_4}}$\Big); Cx Motor\_4],(e413,f413)];  \\
\hspace*{0.2cm}   1,T,[F,F,(F,0),2,[Cx (\&1); Cx (\&1)], momentum\_der e421,inertance\_f Jm\_4 e421 p\_0;   \\
\hspace*{0.2cm}   T,F,(F,0),4,[Cx (\&1); Cx (\&1)],(res\_e Dm\_4 f422,f422); T,T,(F,0),6,[Cx \Big($\dfrac{\texttt{\&1}}{\texttt{Motor\_4}}$\Big);  \\
\hspace*{0.2cm}     Cx Motor\_4],(e423,f423); T,F,(F,0),5,[Cx \Big($\dfrac{\texttt{\&1}}{\texttt{Geer\_4}}$\Big); Cx Geer\_4],(e424,f424)];   \\
\hspace*{0.2cm}    2,F,[F,F,(F,0),2,[Cx (\&1); Cx (\&1)], (momentum\_der e431,inertance\_f J\_4 e431 p\_0);  \\
\hspace*{0.4cm}   F,T,(F,0),5,[Cx Geer\_4; Cx \Big($\dfrac{\texttt{\&1}}{\texttt{Geer\_4}}$\Big)],(e432,f432); T,F,(F,0),0,[Cx (\&1); Cx (\&1)],  \\
\hspace*{1.1cm}    (e433,f433)]; 3,T,[T,F,(F,0),3,[Cx (\&1); Cx (\&1)],  \\
\hspace*{1.5cm}   (compliance\_e K\_2 f441 q\_0, displacement\_der f441 q\_0); F,T,(F,0),0,   \\
\hspace*{0.2cm}   [Cx (\&1); Cx (\&1)],(e442,f442); T,F,(F,0),4,[Cx (\&1); Cx(\&1)],(res\_e D\_4 f443,f443)]]
}}}}
\end{definition}

\noindent Where \texttt{\textsf{D\_4,Ra\_4,Dm\_4}} represent the damping of pulleys, resistance and damping of the motor, respectively. The complex-valued constants \texttt{\textsf{Jm\_4,J\_4}} represents inertial mass of motor and frames, pulleys and strings, respectively. \texttt{\textsf{Motor\_4,Geer\_4}} is the ratio of the gyrator and gear, respectively~\cite{saeed2019comprehensive,formalizationbg}.\\
Similarly, we formalized thumb up and down movements  of the prosthetic hand~\cite{formalizationbg}.
The next step is to formalize the state-space model of the index finger (flex. and ext.) that requires system and input matrices, state and input vectors, and derivative of the state vector, as given in Definition~\ref{DEF:ss_model}. We only provide the formal model of the system matrix here. The formalization of the other vectors and matrices can be found at our project webpage~\cite{formalizationbg}. The system matrix required for the state-space representation of index finger movements (flex.and ext) is defined in \holl~as follows:

\begin{definition}
\label{DEF:sys_mat}
{
\textup{\texttt{\textsf{
$\vdash_{}$  $\forall$$\mathtt{Ra_4\ Motor_4\  Jm_4\ Dm_4\ Geer_4\ J_4\ D_4\ K_2.\ index\_finger\_flex\_ext\_}$    \\
\hspace*{0.4cm} $\mathtt{\mathsf{system\_mat \ Ra_4\ Motor_4\  Jm_4\ Dm_4\ Geer_4\ J_4\ D_4\ K_2\ = \ \begin{bmatrix} \texttt{\textsf{a11\ \ \ a12\ \ \ a13}}\\ \texttt{\textsf{a21\ \ \ a22\ \ \ a23}}\\ \texttt{\textsf{a31\ \ \ a32\ \ \ a33}} \end{bmatrix}}}$
}}}}
\end{definition}

\noindent Where
\small{\texttt{\textsf{a11}} = \texttt{\textsf{Cx \Big(\Big($\dfrac{\mathtt{- Dm_4}}{\mathtt{Jm_4}}$\Big) - \Big($\dfrac{\mathtt{(Geer_4) \ pow 2 \ast D_4}}{\mathtt{Jm_4}}$\Big) - \Big($\dfrac{\mathtt{(Motor_4) \ pow 2}}{\mathtt{Jm_4 \ast Ra_4}}$\Big)\Big)}} \\
\texttt{\textsf{a12}} = \texttt{\textsf{Cx \Big($\dfrac{\mathtt{\mathsf{Geer_4 \ast D_4}}}{\mathtt{\mathsf{J_4}}}$\Big)}} ,
\texttt{\textsf{a13}} = \texttt{\textsf{Cx \Big($\dfrac{\mathtt{\mathsf{- Geer_4}}}{\mathtt{\mathsf{K_2}}}$\Big)}} ,
\texttt{\textsf{a21}} = \texttt{\textsf{Cx \Big($\dfrac{\mathtt{\mathsf{Geer_4 \ast D_4}}}{\mathtt{\mathsf{Jm_4}}}$\Big)}}, \texttt{\textsf{a22}} = \texttt{\textsf{Cx \Big($\dfrac{\mathtt{\mathsf{- D_4}}}{\mathtt{\mathsf{J_4}}}$\Big)}},
\texttt{\textsf{a23}} = \texttt{\textsf{Cx \Big($\dfrac{\texttt{\textsf{\&1}}}{\mathtt{\mathsf{K_2}}}$\Big)}},
\texttt{\textsf{a31}} = \texttt{\textsf{Cx \Big($\dfrac{\mathtt{\mathsf{Geer_4}}}{\mathtt{\mathsf{Jm_4}}}$\Big)}}, \texttt{\textsf{a32}} = \texttt{\textsf{Cx \Big($\dfrac{\texttt{\textsf{- \&1}}}{\mathtt{\mathsf{J_4}}}$\Big)}},
\texttt{\textsf{a33}} = \texttt{\textsf{Cx (\&0)}} }

Next, we formally verify the implementation of the index finger for the flexion and extension movements, connection between index finger and thumb, simplification of the index finger equations and state-space representation of the index finger. These verification results can be found at our project webpage. \\
Finally, we verify the stability of the index finger (flex. and ext.) by utilizing the formally verified system matrix \texttt{\textsf{index\_finger\_flex\_ext\_system\_mat}} in~\holl~as the following theorem:

\begin{theorem}
\label{THM:index_fin_flex_stable}
{
\textup{\texttt{\textsf{
$\vdash_{}$  $\forall$Ra\_4 Motor\_4 Jm\_4 Dm\_4 Geer\_4 J\_4 D\_4 K\_2.   \\
\textbf{[A1]}  Cx \Big($\dfrac{\mathtt{D_4}}{\mathtt{J_4}}$ -  \Big($\dfrac{\mathtt{- Dm_4}}{\mathtt{Jm_4}}$ - $\dfrac{\mathtt{(Geer_4)^2 \ast D_4}}{\mathtt{Jm_4}}$ - $\dfrac{\mathtt{(Motor_4)^2}}{\mathtt{Jm_4 \ast Ra_4}}$\Big)\Big) = Cx b1 +  Cx r $\land$   \\
\textbf{[A2]}  Cx \Big($\dfrac{\mathtt{Geer_4}}{\mathtt{K_2}}$ $\ast$ $\dfrac{\mathtt{Geer_4}}{\mathtt{Jm_4}}$ - $\dfrac{\mathtt{Geer_4 \ast D_4}}{\mathtt{J_4}}$ $\ast$ $\dfrac{\mathtt{Geer_4 \ast D_4}}{\mathtt{Jm_4}}$ - $\dfrac{\texttt{\&1}}{\mathtt{K_2}}$ $\ast$ $\dfrac{\texttt{- \&1}}{\mathtt{J_4}}$ +   \\
 \Big($\dfrac{\mathtt{- Dm_4}}{\mathtt{Jm_4}}$ - $\dfrac{\mathtt{(Geer_4)^2 \ast D_4}}{\mathtt{Jm_4}}$ - $\dfrac{\mathtt{(Motor_4)^2}}{\mathtt{Jm_4 \ast Ra_4}}$\Big) $\ast$ $\dfrac{\mathtt{- D_4}}{\mathtt{J_4}}$\Big) = Cx c1 + Cx (b1 $\ast$ r) $\land$     \\
\textbf{[A3]}  Cx \Big(- $\dfrac{\mathtt{Geer_4 \ast D_4}}{\mathtt{J_4}}$ $\ast$ $\dfrac{\mathtt{Geer_4}}{\mathtt{Jm_4}}$ $\ast$ $\dfrac{\texttt{\&1}}{\mathtt{K_2}}$ - $\dfrac{\mathtt{- Geer_4}}{\mathtt{K_2}}$ $\ast$ $\dfrac{\mathtt{Geer_4 \ast D_4}}{\mathtt{Jm_4}}$ $\ast$ $\dfrac{\texttt{- \&1}}{\mathtt{J_4}}$ +      \\
\hspace*{0.1cm}   \Big($\dfrac{\mathtt{- Dm_4}}{\mathtt{Jm_4}}$ - $\dfrac{\mathtt{(Geer_4)^2 \ast D_4}}{\mathtt{Jm_4}}$ - $\dfrac{\mathtt{(Motor_4)^2}}{\mathtt{Jm_4 \ast Ra_4}}$\Big) $\ast$$\dfrac{\texttt{\&1}}{\mathtt{K_2}}$ $\ast$ $\dfrac{\texttt{- \&1}}{\mathtt{J_4}}$ + $\dfrac{\mathtt{- Geer_4}}{\mathtt{K_2}}$ $\ast$
     $\dfrac{\mathtt{Geer_4}}{\mathtt{Jm_4}}$ $\ast$ $\dfrac{\mathtt{- D_4}}{\mathtt{J_4}}$\Big) = Cx (c1 $\ast$ r) $\land$        \\
\textbf{[A4]}   \big(\&0 < r $\lor$ \big(\&0 < b1 $\land$ \big(b1 pow 2 - \&4 $\ast$ c1 < \&0 $\lor$ b1 pow 2 - \&4 $\ast$ c1 = \&0\big) $ \lor $    \\
\hspace*{0.8cm}   \big(\&0 < b1 pow 2 - \&4 $\ast$ c1 $\land$ \big((b1 pow 2 - \&4 $\ast$ c1) < b1 $\lor$    \\
\hspace*{1cm}   -- b1 < $\sqrt {\texttt{(b1 pow 2 - \&4 $\ast$ c1)}}$ \big)\big)\big)\big) $\Rightarrow$    \\
\hspace*{-0.2cm}    stable\_sys (index\_finger\_flex\_ext\_system\_mat Ra\_4 Motor\_4 Jm\_4 Dm\_4 Geer\_4 J\_4 D\_4  K\_2)
}}}}
\end{theorem}

Assumptions \texttt{\textsf{A1-A3}} capture the necessary conditions for the stability of the index finger (flex. and ext.). Finally, the conclusion provides the stable index finger. The verification of Theorem~\ref{THM:index_fin_flex_stable} is mainly based on Theorem~\ref{THM:stable_mat} alongwith some complex arithmetic reasoning.

\begin{figure}[!ht]
\centering
\resizebox{0.53\hsize}{0.30\textheight}{\includegraphics{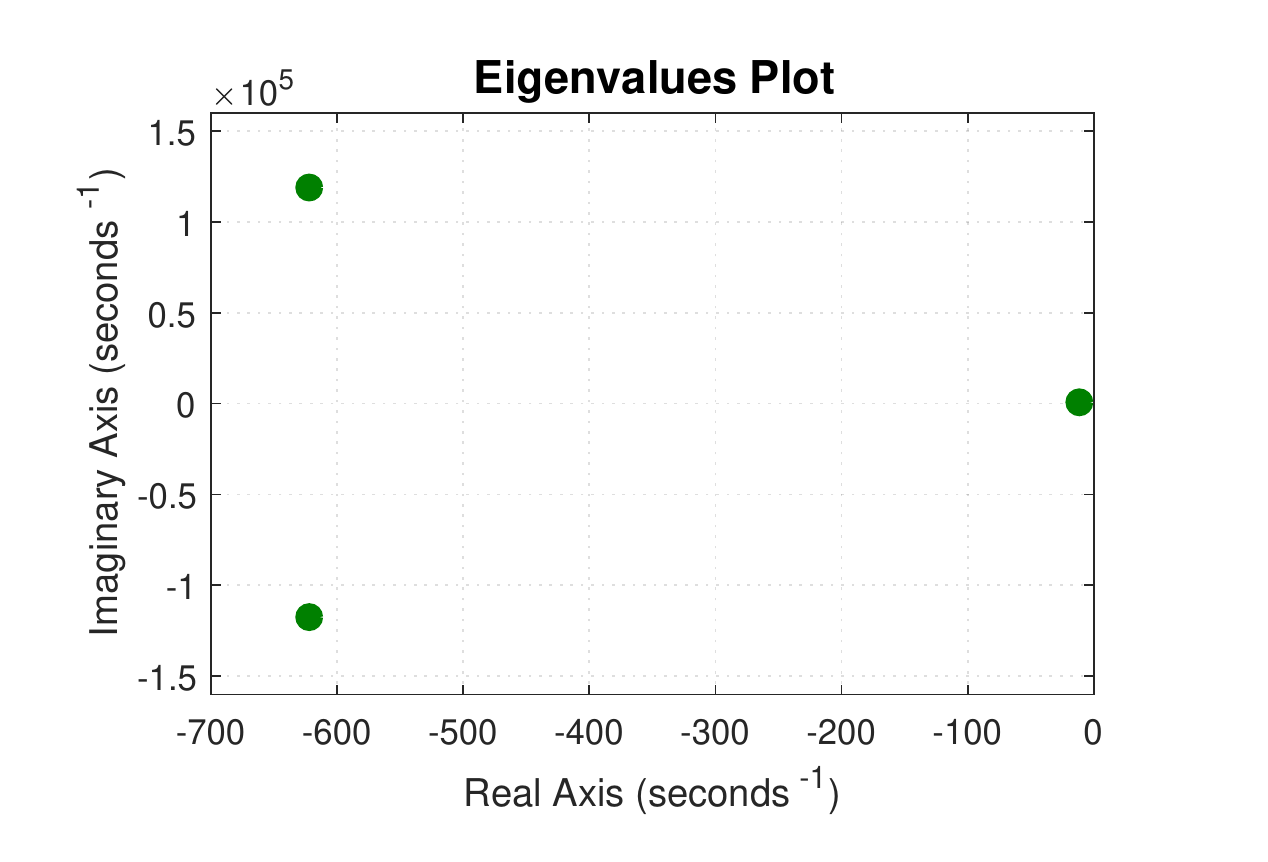}}
\caption{Stable Index Finger (Flex. and Ext.)}
\label{fig:index_fin_matlab}
\end{figure}

To verify the assumptions for all possible values of the parameters, we encoded our formalized Theorem~\ref{THM:stable_mat} in MATLAB in the form of an algorithm to verify the status (stable, marginally stable or unstable) of the given system and it provides results in the graphical form as shown in the Figure~\ref{fig:index_fin_matlab}.
We provided the values of the parameters such as \texttt{\textsf{Ra\_4, Motor\_4, Jm\_4, Dm\_4, Geer\_4, J\_4, D\_4}} and \texttt{\textsf{K\_2}} as an input to the MATLAB algorithm and it provides a graph of the eigenvalues as an output. All of the roots of a polynomial (eigenvalues) lie in the left half of the complex plane in the Figure~\ref{fig:index_fin_matlab}, which indicates the system of index finger (flex. and ext.) is stable.
\par
The distinguishing features of our proposed framework, as compared to the traditional analysis techniques, include $1$) all of the parameters of BG are formally (clearly) defined along with a data type, $2$) all of the variables and functions are of generic nature, i.e., universally quantified and $3$) the utilization of the formally verified algorithm of BG. Soundness is assured in functions and theorems as every new theorem is verified by applying basic axioms and inference rules or any other previously verified theorems/inference rules due to the usage of an interactive theorem prover. In the traditional analysis techniques, like computer-based simulations, there are chances of errors due to the misinterpretation of the parameters and their sampling based nature. We have demonstrated the effectiveness of our formal stability analysis approach by analyzing an anthropomorphic prosthetic mechatronic hand using \holl. It can be seen in our final results that all of the verified theorems are of generic nature, i.e., all of the functions and variables are universally quantified and thus can be specialized based on the requirement of the analysis of the anthropomorphic prosthetic mechatronic hand. Whereas, in the case of computer based simulations, i.e., the MATLAB based analysis, we need to model each case individually. Also, the inherent soundness of the theorem proving technique ensures that all the required assumptions are explicitly present along with the theorem, which are generally ignore in the corresponding MATLAB based analysis and their absence may affect the accuracy of the corresponding analysis. Therefore, we encoded our formally verified stability theorems alongside all of its assumptions in MATLAB, which certify the analysis performed in MATLAB. Moreover, due to the undecidable nature of the higher-order logic, the proposed formalization involves a significant user involvement. However, we have developed simplifiers, such as, \texttt{\textsf{PATHS\_SELECTION\_TAC, BACKWRD\_PATH\_TAC, FWRD\_PATH\_TAC, CASE\_SELECTION\_TAC}} that significantly reduce the user guidance in the proof reasoning process. Finally, the verification and the formal analysis of a BG representation ensures the accurate results and stability of a dynamic system (Anthropomorphic Robotic Hand).


\section{Conclusion}\label{SEC:Conclusion}

In this paper, we proposed a framework for formally analyzing BG representations of engineering and physical systems in the~\holl~theorem prover. Firstly, we proposed an analysis methodology of BG that contains components, laws, causal paths, causal loop, branch and state-space model etc. Secondly, we formally verified all the concepts/steps of BG presented in its analysis methodology. Thirdly, the theorems of generic nature are verified for the stability analysis of matrices. This formal stability analysis requires the formalization of the complex matrices, which are also developed as a part of the proposed framework. Moreover, we encoded these formally verified theorems of matrices in MATLAB to perform the formal stability analysis. Finally, we performed the stability analysis of an anthropomorphic mechatronic prosthetic hand to obtain the accurate state-space model and stable movements performed by different fingers of a robotic hand.
In future, our aim is to extend our current formalization of BGs to incorporate algebraic loops and non-linear systems. In addition, we plan to apply our formalization of BGs to some other complex applications involving robotics and biological systems.

\bibliographystyle{spmpsci}
\bibliography{bibliotex}

\end{document}